\documentclass{aa}
\usepackage{graphicx}

\usepackage{txfonts}

\usepackage{natbib}        
\bibpunct{(}{)}{;}{a}{}{,} 
\usepackage{tabularx}

\begin{document}
   \title{Encounters in the ONC - \\
   observing imprints of star-disc interactions}

   \titlerunning{Encounters in the ONC}

   \author{C. Olczak\inst{1}
     \and S. Pfalzner\inst{1}
     \and A. Eckart\inst{1,2}}

   \institute{I. Physikalisches Institut, Universit\"{a}t zu K\"{o}ln, Z\"{u}lpicher Str.77, 50937 K\"{o}ln, Germany \\
     \email{olczak@ph1.uni-koeln.de}
     \and Max-Planck-Institut f\"{u}r Radioastronomie, Auf dem H\"{u}gel 69, 53121 Bonn, Germany}

   \date{Received ; accepted}

%
%

  \abstract
   {The external destruction of protoplanetary discs in a clustered environment acts mainly due to two mechanisms: gravitational drag by stellar
     encounters and evaporation by strong stellar winds and radiation. It is a fundamental question whether either of these mechanisms is important for the
     stellar evolution and planet formation process in young star clusters.}
   {We focus on the effect of stellar encounters in young dense clusters and investigate whether there are any observables that could trace this
     mechanism and its impact on disc evolution. If encounters play a role in disc destruction, one would expect that stars devoid of disc material would
     show unexpectedly high velocities as an outcome of close interactions. We want to quantify this effect by numerical simulations and compare it to observations.}
   {As a model cluster we chose the Orion Nebula Cluster (ONC). We reanalyzed observational data of the ONC to find encounter signatures in
     the velocity distribution and a possible correlation with signatures of circumstellar discs. We use the
     {\textsc{\mbox{nbody6\raise.2ex\hbox{\tiny{++}}}}} code to model the dynamics of an ONC-like cluster and analyze the velocity distribution and the
     disc-mass loss due to encounters.}
   {We found from the observational data that 8 to 18 stars leave the ONC with velocities several times the velocity dispersion. The majority of these
     high-velocity stars are young low-mass stars ($t$\,$\lesssim$\,$10^5$\,yr, $m$\,$\approx$\,0.2-0.3\,{\mbox{M$_{\odot}$}}), among them several lacking infrared excess
     emission. Interestingly, the high-velocity stars are found only in two separate regions of the ONC - i) close to the cluster centre and ii) in the
     outer cluster region. Our simulations give an explanation for the location of the high-velocity stars and provide evidence for a strong correlation
     between location and disc destruction.}
   {The high-velocity stars can be explained as the outcome of close three-body encounters; the partial lack of disc signatures is attributed to
     gravitational interaction. The spatial distribution of the high-velocity stars reflects the initial structure and dynamics of the ONC. 
     Our approach can be generalized to study the evolution of other young dense star clusters, like the Arches cluster, back in time.
   }

   \keywords{stellar dynamics -
             methods: N-body simulations, observational -
             stars: pre-main sequence, circumstellar matter}

   \maketitle

%

\section{Introduction}

According to current knowledge, planetary systems form from the accretion discs around young stars. These young stars are in most cases not isolated but
are part of a cluster \citep[e.g.][]{2003ARA&A..41...57L}. Densities in these cluster environments vary considerably, spanning a range of 10\,pc$^{-3}$
(e.g. $\eta$ Chameleontis) to $10^6$\,pc$^{-3}$ (e.g. Arches Cluster) . It is still an open question as to how far interactions with the surrounding
stars influence planet formation in dense young clusters ($n$\,$\gtrsim$\,$10^4$\,pc$^{-3}$). These discs disperse over time
\citep{2001ApJ...553L.153H,2002astro.ph.10520H,2006ApJ...638..897S,2008ApJ...672..558C} and in dense clusters the disc frequency seems to be lower
in the core \citep[e.g.][]{2007ApJ...660.1532B}. This is attributed to external violent processes like
photo-evaporation or encounter-induced disc mass loss, mainly caused by the massive stars that are concentrated in the cluster core.

Earlier numerical investigations seemed to indicate that photoevaporation should by far dominate the external disc destruction
\citep{2001MNRAS.325..449S,2004ApJ...611..360A}. The question of whether encounters play a vital role in the formation process of stars and planets has
 been studied far less and is still open \citep[e.g.][]{2006ApJ...641..504A}. Only recently has it been shown that stellar encounters do have an effect
 on the discs surrounding stars in a young dense cluster
 \citetext{\citealp[][]{2006ApJ...642.1140O,2006A&A...454..811P,2006ApJ...652L.129P,2006ApJ...653..437M,
     2007ApJ...661L.183M,2007ApJ...656..275M,2007A&A...462..193P,2007A&A...475..875P}; \citealp[see also the review by][]{2007ARA&A..45..481Z}}. The
 importance of encounters has been underestimated because in previous studies the focus has been on encounters between solar mass stars
 \citep[e.g.][]{1991MNRAS.249..584C,1995ApJ...455..252H}. However, discs are most affected when the masses of the stars involved in the encounter are
 unequal \citep{2006ApJ...642.1140O,2007ApJ...656..275M}. Moreover, the massive stars in the center of a stellar cluster act as gravitational foci for
 the lower mass stars and are thus subject to repeated encounters. These accumulated perturbations can lead to a total destruction of a massive star's
 disc \citep{2006A&A...454..811P}.

The main focus of this paper is the question of whether there is direct \emph{observational} evidence for encounters among star-disc systems in young clusters
and for disc-mass loss due to encounters. It is difficult to distinguish observationally whether photoevaporation or gravitational
interaction are responsible for the loss of (outer) discs. The reason is that in both cases interaction with a massive star is the most destructive
process. Thus the observation of a decreased disc frequency in the cluster core, as mentioned above, does not allow one to favor either of the two
mechanisms. The difficulty of tracing stellar encounters directly by observations is their short duration. Direct imaging of tidal tails of a disc would
be the best proof but the probability of such an event is very low: the perturbed disc circularizes and tidal tails dissipate on a time scale of
$\lesssim$\,1000\,yr. Nevertheless, there exist observations of tidal tails or spiral arms in star-disc systems
\citep[e.g.][]{2005prpl.conf.8092B,2006A&A...452..897C,2006ApJ...645.1297L}, but often it is unclear whether these can be attributed to the passage of an
unbound perturber or a binary companion or gravitational instabilities caused by a giant planet. Due to the short dissipation time scale of
encounter-induced tidal tails, there would be a high probability for the encounter partner still to be located close ($\lesssim$\,$10^4$\,AU) to the
perturbed disc. However, source confusion in dense young clusters like the ONC, where stellar encounters are frequent enough to study the effect on
stellar discs, would prevent a clear identification. Set in relation to the rate of encounters in which prominent tidal tails may be formed, we estimate
an upper limit of just four candidate stars for a direct observation of tidal tails in a protoplanetary disc due to encounters in the ONC (see Appendix
\ref{app:enc_rate} for a detailed calculation). However, an unambiguous imprint of an encounter among stars is the high velocity of a star which has been
expelled in a close gravitational interaction, mostly as a result of a three-body encounter \citep[see][]{1975MNRAS.173..729H}. Thus an analysis of the
velocity distribution of a cluster is the key to finding candidates of close encounters between young stars.

Here the ONC is used as a prototype young cluster because it is one of the best-studied regions in our galaxy, and the only young dense cluster for which
velocities of more than 1000 of its members have been determined \citep{1988AJ.....95.1755J}. In addition, its high density suggests that stellar
encounters might be relevant for the evolution of circumstellar discs. Throughout this work we assume that initially all stars are surrounded by
protoplanetary discs. This is justified by observations that reveal disc fractions of nearly 100\,\% in very young star clusters
\citep[e.g.][]{2000AJ....120.1396H,2000AJ....120.3162L,2001ApJ...553L.153H,2005astro.ph.11083H}.

In \S\ref{sec:obs} we present results from a search for candidate stars of close encounters in the publicly available observational data of the ONC. For
this purpose we have reanalyzed the investigations of \citet{1988AJ.....95.1755J}, \citet{1997AJ....113.1733H}, and \citet{1998AJ....116.1816H} for
stellar velocities and infrared excess. The
basic properties of the ONC used for our numerical model are described in \S\ref{sec:onc_dyn}. There we also discuss theoretical estimates and observational data related to
binary populations and encounters. Afterwards we present results from a
numerical approach to this problem (\S\ref{sec:sim}). Observational and numerical results are summerized and discussed in \S\ref{sec:discussion}.

%

\section{Observational data}

\label{sec:obs}

Our search for candidate stars of close encounters from observational data concentrates on velocity surveys of Orion. There exists only one large data
set of the ONC that provides velocity information for several hundred stars, the proper motion survey of \citet{1988AJ.....95.1755J}. Fortunately, the
same stars have been analyzed by \citet{1998AJ....116.1816H} for infrared excess emission as an indicator of circumstellar material. They defined
a quantity measuring the magnitude of the near-infrared excess, $\Delta(I_{\mathrm{C}}-K)$,
\begin{eqnarray}
\Delta(I_{\mathrm{C}}-K) = (I_{\mathrm{C}}-K)_{\mathrm{observed}} - 0.5A_V - (I_{\mathrm{C}}-K)_{\mathrm{photosphere}},
\end{eqnarray}
where the first term is the observed $I_{\mathrm{C}}-K$ color, the second term the contribution of reddening calculated from the extinction values derived
from $V-I_{\mathrm{C}}$ colors as in Hillenbrand (1997), and the third term the contribution of the underlying stellar photosphere. This measure of
the near-infrared excess, $\Delta(I_{\mathrm{C}}-K)$, traces only the innermost ($<$\,0.1\,AU) part of the circumstellar disc. Hence, the absence of such
emission is not to be confused with a complete absence of a protoplanetary disc, yet is dependent on a number of parameters, such as disc accretion rate,
inclination, inner hole, and stellar mass and radius \citep{1998AJ....116.1816H}. However, in the following, we will refer to stars that lack infrared
excess emission as ``discless'', emphasizing that these stars may be well surrounded by circumstellar material but do not show the typical
infrared excess of a young star-disc system.

In addition \citet{1997AJ....113.1733H} has investigated stellar properties like mass and age. The investigation is based on optical photometric and
spectroscopic data and covers only about half of the stellar population of the ONC, while the more embedded stars are not accessible at this
wavelength. However, \citet{1997AJ....113.1733H} states that the investigated stellar sample is representative of the entire population of the
ONC. Masses and ages were derived from an HR diagram via comparison with theoretical pre-main sequence evolutionary tracks. This method leaves some
uncertainty as to the absolute stellar age ($\sim$0.5\,dex, see Appendix~\ref{app:age_errors}) and mass calibrations, with mass
deviations of about factors of two and age differences of several Myr among different models. Additionally, photometric errors translate into uncertainties
of the derived masses, but this is significant only for stars more massive than $M$\,$\approx$\,1.5\,{\mbox{M$_{\odot}$}}. However, all these sources of uncertainties
are less relevant for the present investigation because (i) the stars of interest are of low mass as will be shown later, and (ii) the derived
conclusions are based mainly on relative masses and ages.

We have merged the data of the three investigations described and excluded stars for which measurements of proper motion or infrared excess are missing,
resulting in a database of 655 stars. In order to achieve the most secure distinction possible between stars with infrared excess and pure photospheric
emission, we adopted the criterion of \citet{2005AJ....129..363S} and excluded all stars that have an infrared excess $\Delta(I_{\mathrm{C}}-K)$ in the
range $0<\Delta(I_{\mathrm{C}}-K)<0.5$. A star is classified as discless if $\Delta(I_{\mathrm{C}}-K) \le 0$, or as a star-disc system if
$\Delta(I_{\mathrm{C}}-K) \ge 0.5$. This additional selection criterion reduces the sample of stars that have been used for this investigation to 405,
among them 266 star-disc systems and 139 stars lacking any excess emission.

\begin{figure*}[t]
   \centering
   \includegraphics[height=0.45\linewidth,angle=-90]{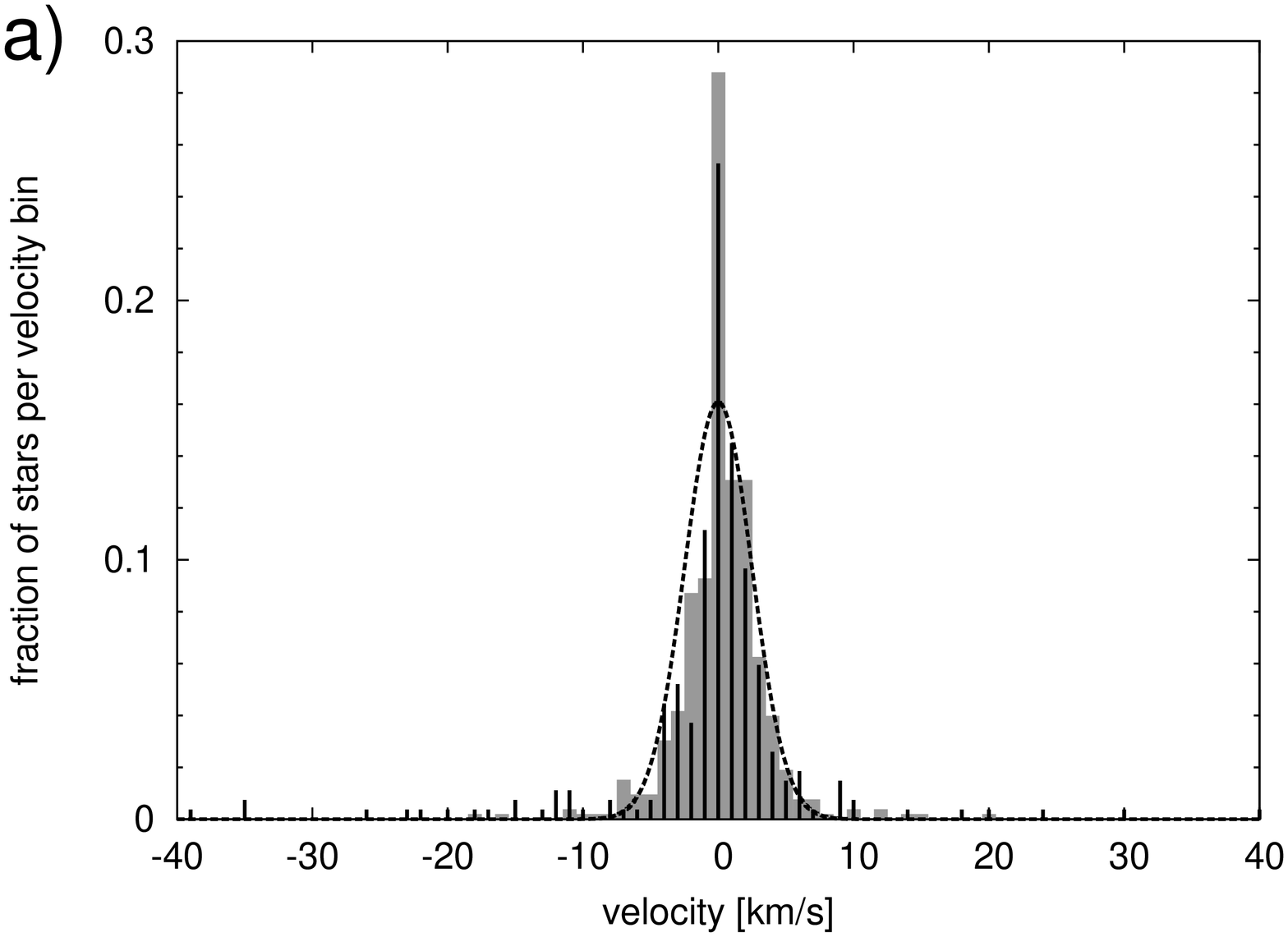}
   \includegraphics[height=0.45\linewidth,angle=-90]{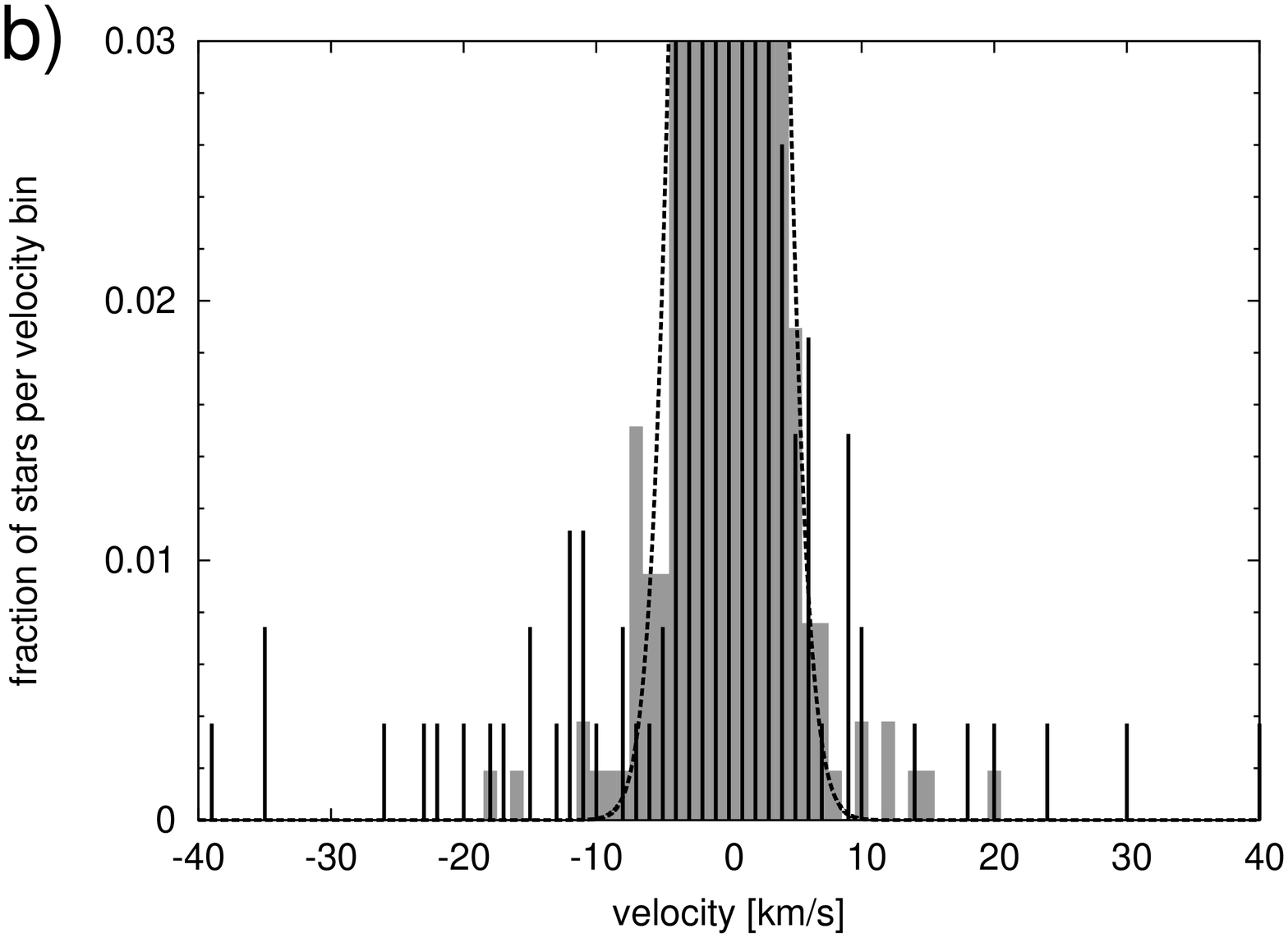}
   \includegraphics[height=0.45\linewidth,angle=-90]{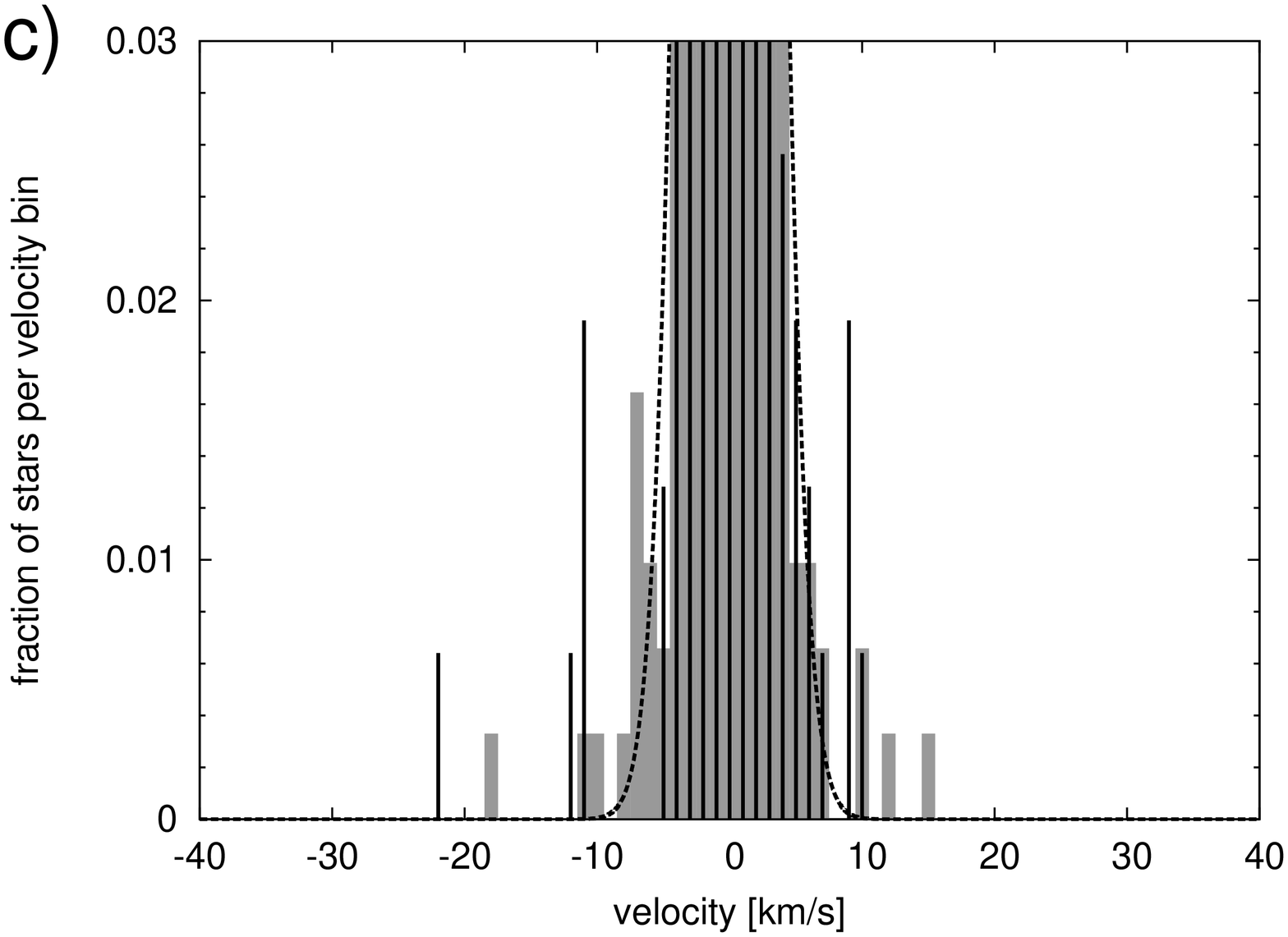}
   \includegraphics[height=0.45\linewidth,angle=-90]{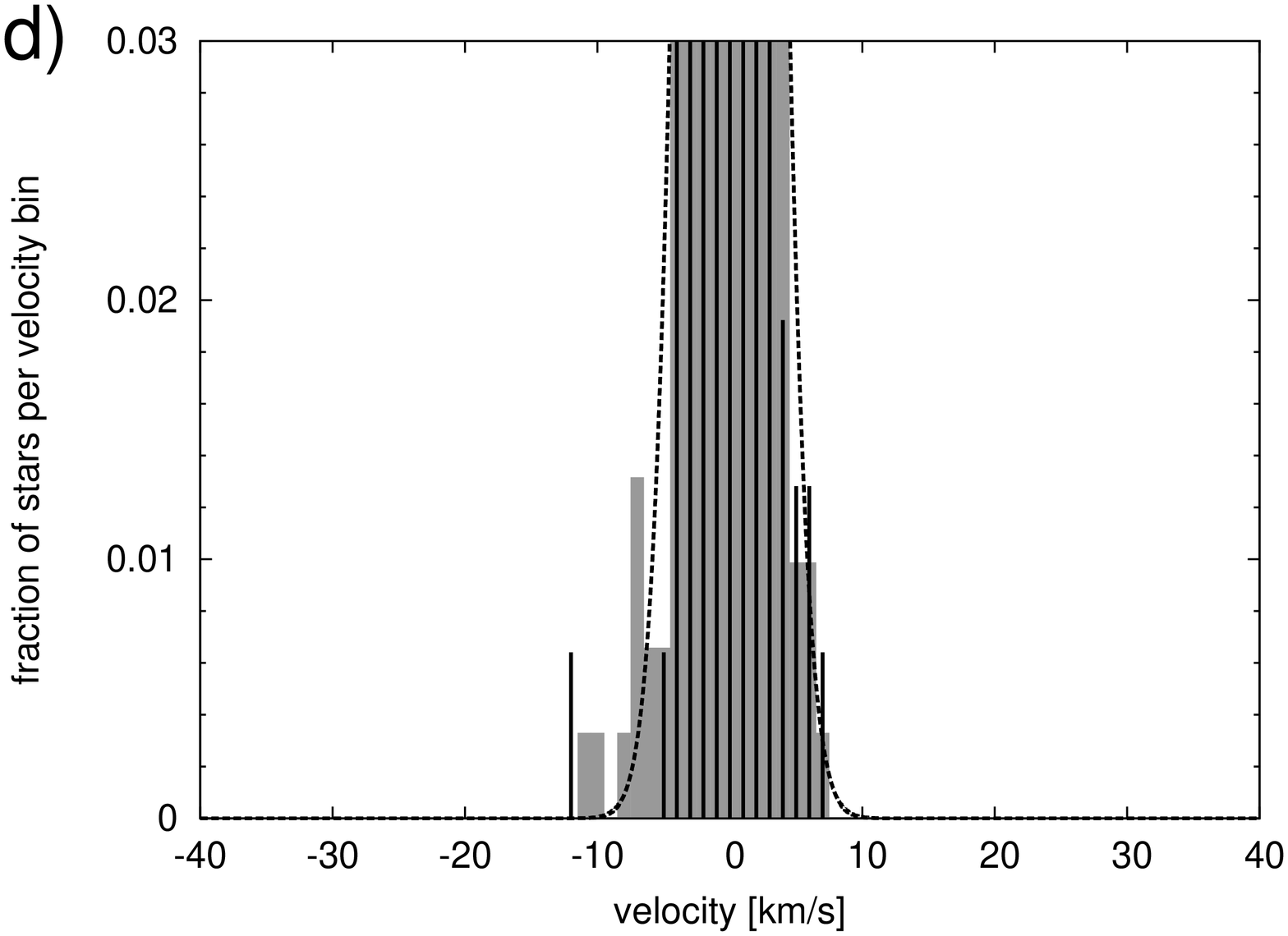}
   \caption{Velocity distribution from observational data, adopting a distance of 420\,pc to the ONC
       \citep[cf.][]{2007MNRAS.376.1109J,2007A&A...474..515M}. See text for details of its construction. The bin width is 0.7\,{\mbox{km\,s$^{-1}$}}. The stars have
     been divided into two groups according to the disc signature $\Delta(I_{\mathrm{C}}-K)$ (see text): grey boxes represent star-disc systems, lines
     represent discless stars. For comparison a Gaussian with dispersion $\sigma$\,=\,{\mbox{$\sigma_{\rm{1D}}^{\rm{JW}}$}}\,=\,2.5\,{\mbox{km\,s$^{-1}$}} \citep{1988AJ.....95.1755J} is superposed
     (dashed line). \textbf{a)} Complete sample of stars in the ONC with available proper motions from \citet{1988AJ.....95.1755J} and infrared excess
     from \citet{1998AJ....116.1816H}. \textbf{b)} Like a), but zoomed into the lower part. \textbf{c)} Like b), but with restricted stellar ages (see
     text). \textbf{d)} Like b), but with restricted stellar ages and velocity errors as described in the text.
   }
   \label{fig:obs_gauss}
\end{figure*}

In Fig.~\ref{fig:obs_gauss} we show the velocity distribution of stars with and without discs, adopting a distance of
  420\,pc to the ONC \citep[cf.][]{2007MNRAS.376.1109J,2007A&A...474..515M}. The velocity distribution has been built by binning \emph{each component} of
the proper motion, $\mu_x$ and $\mu_y$, of each star separately, not its total two-dimensional motion, or in other words by summing the Gaussian velocity
distribution of each spacial direction, $x$ and $y$, which again results in a Gaussian. The reason not to bin the total two-dimensional motion is the
improvement of statistics due to the twice as large data set. For comparison, a Gaussian with a one-dimensional velocity dispersion
$\sigma$\,=\,{\mbox{$\sigma_{\rm{1D}}^{\rm{JW}}$}}\,=\,2.5\,{\mbox{km\,s$^{-1}$}}, as derived by \citet{1988AJ.....95.1755J}, is superimposed (dashed line). The distribution shows the expected
features: At velocities $\mu_{x,y}$\,$\lesssim$\,3{\mbox{$\sigma_{\rm{1D}}^{\rm{JW}}$}} the shape is approximately Gaussian, though a distinct peak at zero velocity is present. This
peak is only prominent for the $x$-components of the proper motion but is independent of the applied binning. We suppose that this is an artifact of
the plate reduction technique used by \citet{1988AJ.....95.1755J}. At higher velocities, $\mu_{x,y}$\,$\gtrsim$\,3{\mbox{$\sigma_{\rm{1D}}^{\rm{JW}}$}}, there is an overabundance of
stars when compared with the theoretically expected Gaussian distribution. In this regime stars are not bound energetically to the cluster
  \citep[see][Eq.~8-3]{1987gady.book.....B}. In the following we will refer to these stars as ``high-velocity stars''.

\begin{figure}[t]
   \centering
   \includegraphics[height=1.0\linewidth,angle=-90]{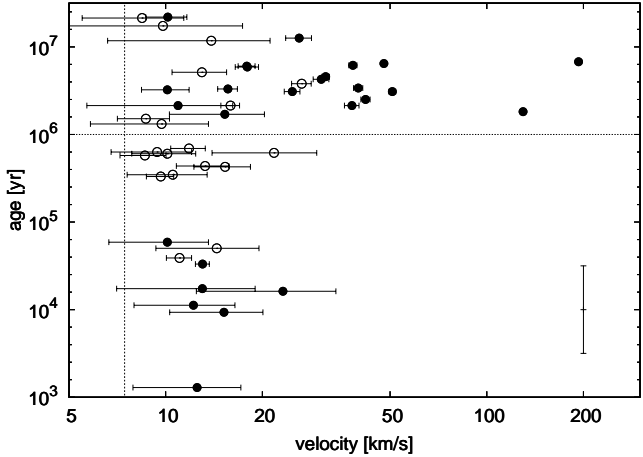}
   \caption{Distribution of age and velocity of the high-velocity stars in the ONC, classified as star-disc systems (open circles) and discless stars
     (filled circles). The mean age of the ONC, $t$\,=\,1\,Myr, and the minimum one-dimensional velocity of the high-velocity sample, 3{\mbox{$\sigma_{\rm{1D}}^{\rm{JW}}$}}, are
     indicated by the horizontal and vertical dashed line, respectively. The uncertainty of the determined ages of $\sim$0.5\,dex (see
       Appendix~\ref{app:age_errors} for the derivation) is indicated by the vertical error bar. At high velocities the (horizontal) velocity error bars are
       smaller than the symbol size.}
   \label{fig:age_vel}
\end{figure}

The number of high-velocity stars, $N_{\rm{hvs,\,obs}}$\,=\,50, is remarkably high, much higher than the theoretically expected
$N_{\rm{hvs}}$\,$\lesssim$\,15, as derived in \S~\ref{sec:onc_enc}. In fact, $N_{\rm{hvs,\,obs}}$ should be even
lower than the theoretical $N_{\rm{hvs}}$ because only about $1/5$ of the ONC stellar population is covered by the observational
data. The reason is most likely a contribution from foreground stars, which show large proper motions due to projection effects. In order to exclude
probable foreground stars from the sample of ONC stars, we plot in Fig.~\ref{fig:age_vel} age vs. velocity for the high-velocity stars, again for the
two groups of discless stars and star-disc systems. Two opposite trends are visible: The discless stars with ages $t$\,$>$\,1\,Myr have a $\sim$10 times higher
maximum velocity than younger ones ($\sim$200\,{\mbox{km\,s$^{-1}$}} compared to $\sim$20\,{\mbox{km\,s$^{-1}$}}) and a twice as high mean velocity ($\sim$13\,{\mbox{km\,s$^{-1}$}} compared to
$\sim$30\,{\mbox{km\,s$^{-1}$}}). This strong correlation is likely due to a large fraction of foreground objects among the stars with ages $t$\,$>$\,1\,Myr, and since this age
boundary is consistent with the mean cluster age, it marks a conservative upper limit to the age of probable cluster members. Conversely, the star-disc
systems are similarly distributed in both age groups, with comparable maximum and median velocities ($\sim$25\,{\mbox{km\,s$^{-1}$}} and $\sim$11\,{\mbox{km\,s$^{-1}$}},
respectively). However, we exclude \emph{all} stars with ages $t$\,$>$\,1\,Myr in the further discussion for three reasons: (i) the age of these stars exceeds
our simulated cluster age, so they are dynamically distinct, (ii) at least some of these stars might not be members of the ONC due to their high age
(and velocity), and (iii) protoplanetary discs with ages $t$\,$>$\,1\,Myr may be subject to significant evolution due to internal
processes{\footnote{According to studies of properties and evolution of protoplanetary discs in young clusters
    \citep[e.g.][]{2005astro.ph.11083H,2006ApJ...638..897S}, it is valid to assume that the protoplanetary discs in the ONC have been only marginally
    subject to internal disc processes. As such, we assume in the following that the effect of external processes on discs, i.e. photevaporation and
    encounters, have not been masked by internal processes at the current age of the ONC.}}, so that the effect of encounters could not be quantified.

Of the group of high-velocity stars younger than 1\,Myr, the ``probable high-velocity cluster members'', about half show disc signatures, the rest does
not. The fact that stars with velocities of several tens of {\mbox{km\,s$^{-1}$}} are not present in this group is in accordance with dynamical estimates: the
probability of both the generation and detection of stars with such high velocities is very low due to the need for a very close approach and
short traverse of the cluster.

The uncertainties in the estimated ages of stars with and without infrared excess emission do not alter the fact that some of the youngest
high-velocity stars lack infrared excess. This is noteworthy because in the early stages of stellar evolution one would expect accretion rates to be high
and circumstellar material to be close enough to the star that significant excess emission can be detected. However, because the excess emission measured by
$\Delta(I_{\mathrm{C}}-K)$ originates close to the stellar surface, it is strongly dependent on the geometry and orientation of the disc. A more robust and sensitive
indicator of circumstellar discs is the $K-L$ color \citep{1997AJ....114..288M,2000AJ....120.1396H}, tracing material out to radii of $\sim$0.1\,AU
\citep{2005ApJ...627L..57H}. We have thus cross-checked the ``discless'' stars with $\Delta(I_{\mathrm{C}}-K) \le 0$ for excess emission at longer
wavelengths. The results are shown in Table~\ref{tab:obs_highvel}.

\begin{table*}[t]
\begin{minipage}{\textwidth}
\begin{center}
\begin{tabular}{l*{13}{r}l}
\hline
ID\footnote{Stellar ID from \citet{1988AJ.....95.1755J}}
& $r$\footnote{Projected distance from {\mbox{$\theta^1{\rm{C}}\:{\rm{Ori}}$}}.}
& $\mu_x$\footnote{Proper motion and associated error.}
& $\sigma_x$
& $\mu_y$
& $\sigma_x$
& $P$\footnote{Probability of cluster membership.}
& $J$\footnote{Apparent magnitude in specified band filter.}
& $H$
& $K$
& $L$
& SpT\footnote{Spectral type.}
& $m$\footnote{Stellar mass.}
& $\log{t}$\footnote{Stellar age.}
& excess\footnote{Detection of excess emission.} \\

[JW]            & [pc]  & [{\mbox{km\,s$^{-1}$}}]  & [{\mbox{km\,s$^{-1}$}}]      & [{\mbox{km\,s$^{-1}$}}]  & [{\mbox{km\,s$^{-1}$}}]     & [\%] & [mag] & [mag] & [mag] & [mag] &       & [{\mbox{M$_{\odot}$}}] & [yr]      &       \\
\hline
\object{JW 19}    & 1.99  & -12.2   &   4.2       &   0.2   &   2.2      & 84   & 12.98 & 12.26 & 11.87 &       & M5.5  & 0.16    & 4.05      & no    \\
\object{JW 45}    & 1.12  & -13.0   &   0.6       &  -0.8   &   0.8      &  0   &  8.83 &  8.28 &  8.05 &  7.89 & K3    & 0.94    & 4.52      & no    \\
\object{JW 505}   & 0.10  &  -4.2   &  11.4       & -22.8   &   8.8      & 93   & 12.33 & 11.65 & 11.18 & 10.65 & M2    & 0.20    & 4.21      & ?     \\
\object{JW 510}   & 0.36  &  -5.0   &   2.8       &   8.8   &   2.4      & 95   & 13.01 & 12.18 & 11.76 & 11.16 & M5    & 0.14    & 4.77      & yes   \\
\object{JW 559}   & 0.19  & -11.6   &   3.2       &   9.8   &   3.8      & 99   & 12.43 & 11.74 & 11.30 & 11.04 & M5.5e & 0.16    & 3.97      & no    \\
\object{JW 569}   & 0.24  & -12.0   &   3.8       &  -3.6   &   3.4      & 92   & 12.47 & 11.58 & 11.00 & 10.14 & M3.5  & 0.09    & 3.11      & yes   \\
\object{JW 616}   & 0.30  &   9.4   &   5.6       &   9.0   &   2.8      & 95   & 13.21 & 12.36 & 11.91 & 11.50 & M3.5  & 0.15    & 4.24      & no    \\
\hline
\end{tabular}
\end{center}
\end{minipage}
\caption{Properties of discless high-velocity stars in the ONC.}
\label{tab:obs_highvel}
\end{table*}

Two of the stars previously determined as ``discless'' do show a typical emission signature of warm circumstellar matter. Four
high-velocity stars remain that lack infrared excess. We cannot determine whether the pure photospheric colors point to the absence of a
circumstellar disc. We will discuss this possibility later.

\begin{figure}[t]
   \centering
   \includegraphics[height=0.8\linewidth,angle=-90]{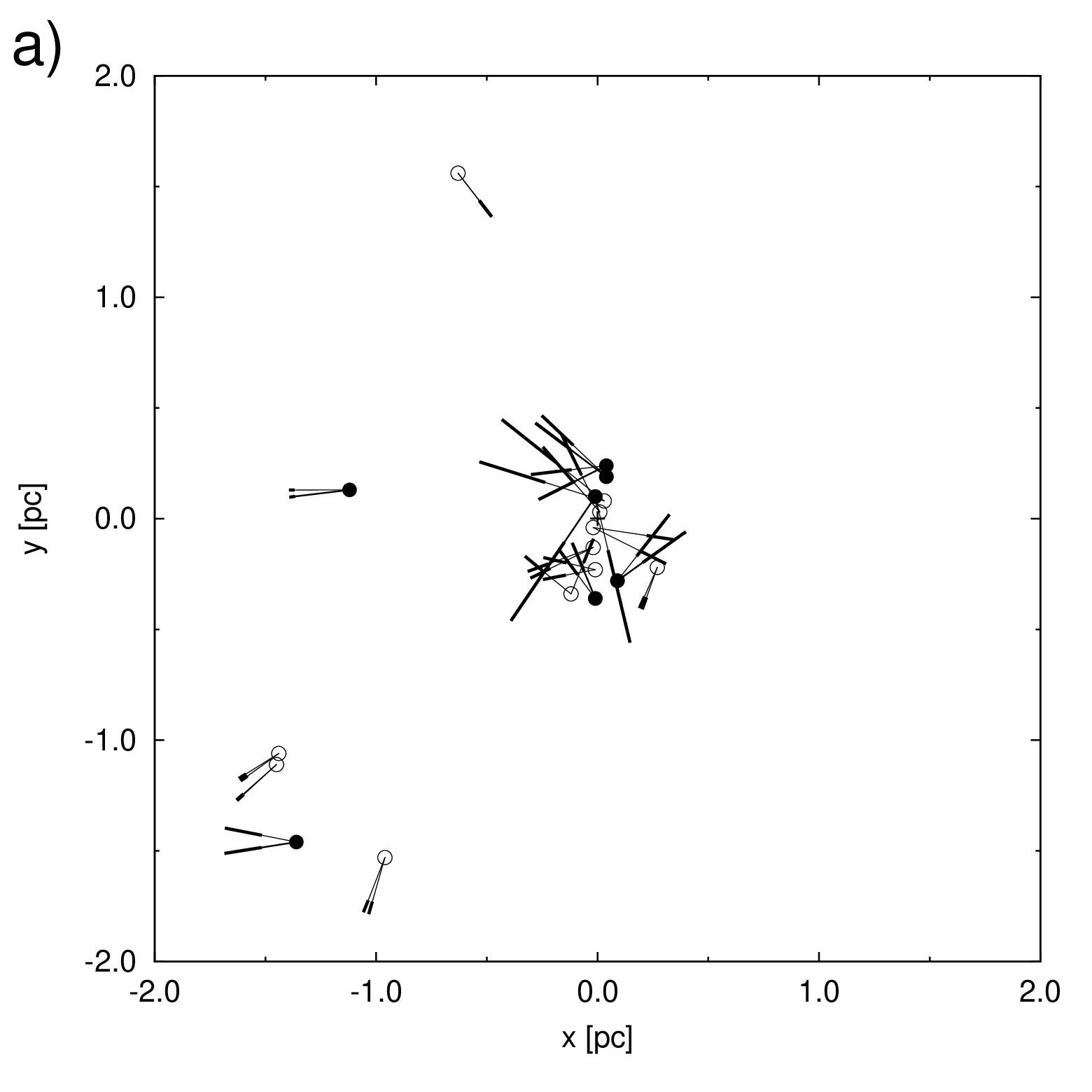}
   \includegraphics[height=0.8\linewidth,angle=-90]{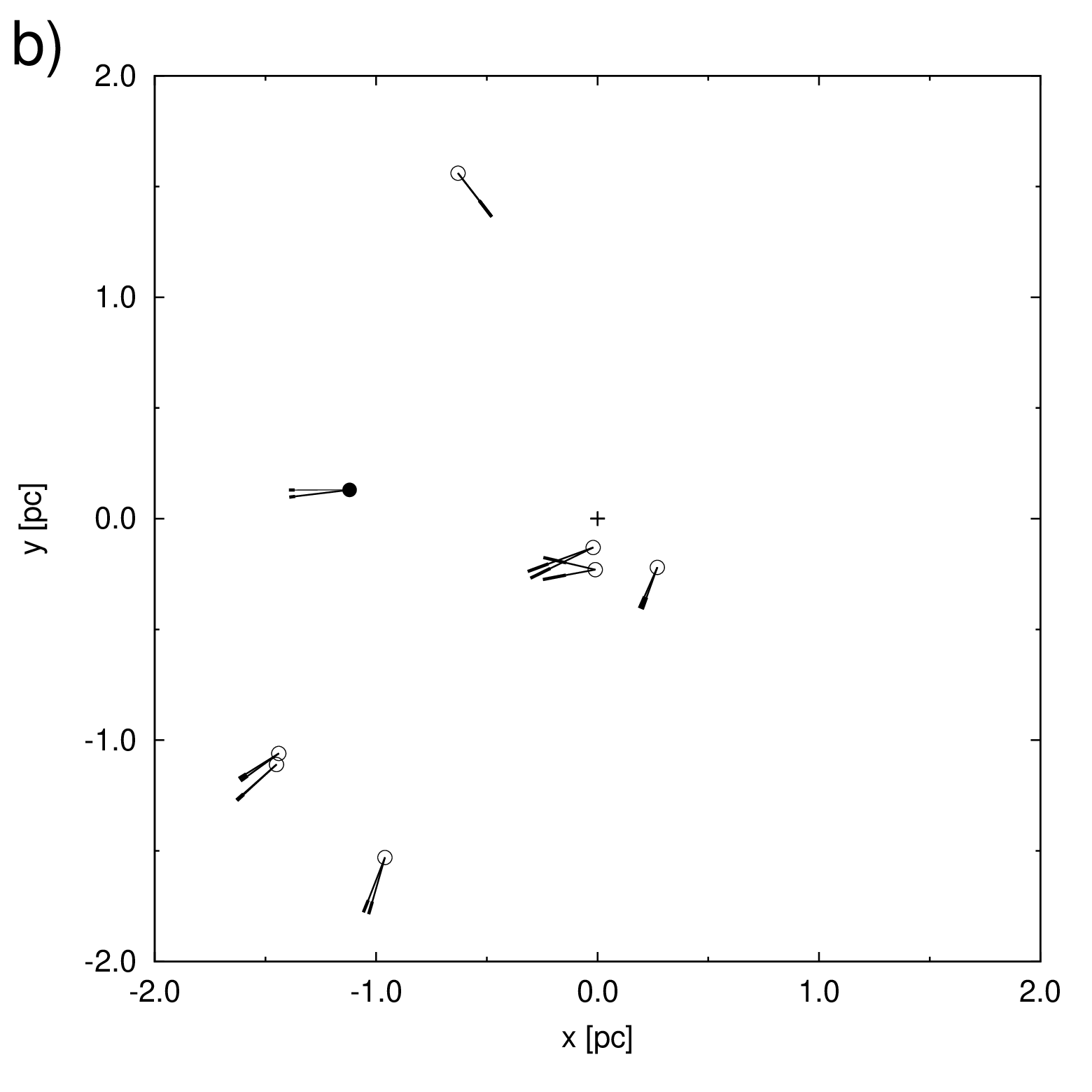}
   \caption{Positions and velocities of high-velocity star-disc systems (open circles) and discless stars (filled circles) in the ONC. The frames are centered on
     the most massive star, {\mbox{$\theta^1{\rm{C}}\:{\rm{Ori}}$}}, which is marked by a cross. Thin lines indicate the distance (in pc) a star would have moved in a period of
     2\,$\cdot$\,$10^4$\,yr. Thick lines and the opening angle reflect the error of the magnitude and the direction of the proper motion,
     respectively. {\textbf{a)}}: Sample of all high-velocity stars with ages $<$\,1\,Myr. {\textbf{b)}}: Restricted sample of high-velocity stars with
     ages $<$\,1\,Myr: only stars fulfilling the constraints on the velocity errors as dicussed in the text are shown.}
   \label{fig:obs_pos_vel}
\end{figure}

We have plotted the positions and velocity vectors of the selected probable high-velocity cluster members in Fig.~\ref{fig:obs_pos_vel}a. Two
features are apparent: (i) Most high-velocity stars are concentrated in the inner tenths of a parsec around the most massive ONC member,
{\mbox{$\theta^1{\rm{C}}\:{\rm{Ori}}$}}, and (ii) ``outliers'' are located more than 1\,pc from the cluster center with some stars moving in radial directions away
from the cluster centre. Although one would expect the former, as most encounters usually happen in the dense cluster centre, there are two
unexpected features: (i) The region between the cluster centre and outer cluster parts, 0.4\,pc\,$\lesssim$\,$r$\,$\lesssim$\,1.0\,pc, is devoid of
high-velocity stars, and (ii) the velocity vectors of several stars in the outer cluster parts do not point away from the cluster centre (as if these
stars had an encounter in the less dense cluster parts). In order to be confident that the encounter candidate stars are indeed high-velocity stars and
not affected by observational uncertainties, we restrict our sample of high-velocity stars to maximum velocity errors of less than 30\,\% or below
2.0\,{\mbox{km\,s$^{-1}$}}. This reduces the number of high-velocity stars to a total of only eight, among them just one star without infrared excess, JW~45. The
  corresponding position-velocity diagram is shown in Fig.~\ref{fig:obs_pos_vel}b (see also the corresponding velocity distribution in
Fig.~\ref{fig:obs_gauss}d). However, apart from the concentration of high-velocity stars close to the cluster centre, the same features as in
Fig.~\ref{fig:obs_pos_vel}a are apparent. The reason for the different central concentration is the strong acceleration of stars close to the cluster
centre which results in large errors of the derived proper motions. Consequently our velocity error criterion preferentially excludes stars close to the
cluster centre. We will address the particular features in \S\ref{sec:sim}, where we present the results from numerical simulations of our dynamical model
of the ONC.

Below we review the known properties of the ONC, which we use to model the cluster in an as realistic a way as possible.

\section{Structure and dynamics of the ONC}

\label{sec:onc_dyn}

The ONC is a rich stellar cluster with about 4000 members with masses $m$\,$\ge$\,0.08\,{\mbox{M$_{\odot}$}} in a volume $\sim$5\,pc across \citep{1998ApJ...492..540H,2000ApJ...540..236H}. Most of the objects are T~Tauri stars. The mean stellar mass is about $\bar{m}$\,$\approx$\,0.5\,{\mbox{M$_{\odot}$}} and the half-mass radius $R_{\rm{hm}}$\,$\approx$\,1\,pc \citep{2005MNRAS.358..742S}. Recent studies of the stellar mass distribution \citep{2000ApJ...540..236H,2000ApJ...540.1016L,2002ApJ...573..366M,2004ApJ...610.1045S} reveal no significant deviation from the generalized IMF of \citet{2002Sci...295...82K},
\begin{eqnarray}
\xi(m)= \left\{
\begin{array}{lrll}
m^{-1.3} & , & 0.08 \le m/{\mbox{M$_{\odot}$}} <0.50, \\
m^{-2.3} & , & 0.50 \le m/{\mbox{M$_{\odot}$}} <1.00, \\
m^{-2.3} & , & 1.00 \le m/{\mbox{M$_{\odot}$}} <\infty.
\end{array} \right.
\label{eqn:imf}
\end{eqnarray}
The shape of the system is not perfectly spherical but elongated in the north-south direction. The probable reason for this asymmetry is the
gravitational potential of a massive molecular ridge in the background of the cluster, OMC~1 \citep{1998ApJ...492..540H}. The mean age of the whole cluster has been estimated to be $t_{\rm{ONC}}$\,$\approx$\,1\,Myr, although a
significant age spread of the individual stars is evident \citep{1997AJ....113.1733H,2000ApJ...540..255P}. Radio observations by
\citet{1997A&A...327.1177W} show that only a few solar masses of ionized gas are present in the inner $\sim$1\,pc.

The density and velocity distribution of the ONC resembles an isothermal sphere. The central number density $\rho_{\rm{core}}$ in the inner 0.053\,pc
reaches $4.7$\,$\cdot$\,$10^4$\,pc$^{-3}$ \citep{2002Msngr.109...28M} and makes the ONC the densest nearby ($<$\,1\,kpc) young stellar cluster. The dense
inner part of the ONC, also known as the Trapezium Cluster (TC), is characterized by $R_{\rm{TC}}$\,$\lesssim$\,0.3\,pc and $N_{\rm{TC}}$\,$\approx$\,750, or $n_{\rm{TC}}$\,$\approx$\,$10^3$\,pc$^{-3}$. In their proper motion study of the ONC, \citet{1988AJ.....95.1755J} found the velocity
dispersion to be nearly constant at all cluster radii and obtained a one-dimensional velocity dispersion {\mbox{$\sigma_{\rm{1D}}^{\rm{JW}}$}}\,=\,2.5\,{\mbox{km\,s$^{-1}$}}. This translates into a
three-dimensional velocity dispersion of {\mbox{$\sigma_{\rm{3D}}^{\rm{JW}}$}}\,=\,$\sqrt{3}$\,{\mbox{$\sigma_{\rm{1D}}^{\rm{JW}}$}}\,=\,4.3\,{\mbox{km\,s$^{-1}$}}. Recently, \citet{2007arXiv0711.0391F} obtained a somewhat higher
one-dimensional velocity dispersion
of $\sigma_{\rm{1D}}^{\rm{F+}}$\,=\,3.1\,{\mbox{km\,s$^{-1}$}} from radial velocity measurements.
However, they caution that their velocity distribution of ONC stars has a peak that is too low compared to the expected Gaussian distribution with dispersion
  $\sigma_{\rm{1D}}^{\rm{F+}}$, so we will rely on the result of \citet{1988AJ.....95.1755J}.

Using this value, the virial ratio $Q_{\rm{vir}}$ of the ONC becomes
\begin{equation}
Q_{\rm{vir}}=\frac{R_{\rm{hm}}\left({\mbox{$\sigma_{\rm{3D}}^{\rm{JW}}$}}\right)^2}{2GM} \approx 1.5, \label{eqn:vir}
\end{equation}
where $M$\,=\,$\bar{m}N$\,$\approx$\,2000\,{\mbox{M$_{\odot}$}}. This indicates that the ONC seems to be gravitationally
unbound ($Q_{\rm{vir}}$\,$>$\,1). However, the gas mass of the background molecular cloud OMC~1 should be
considered. \citet{2006ApJ...644..355H} suggest that the molecular cloud that formed the ONC contained about 6000\,{\mbox{M$_{\odot}$}}. The remaining part of it in the cluster
background still has a mass of about 2000\,{\mbox{M$_{\odot}$}} \citep{1998AJ....116.1816H}.

In the most recent study on circumstellar discs in the Trapezium Cluster, \citet{2000AJ....120.3162L} found a fraction of 80-85\,\% discs among the
stellar population from the $L$-band excess. This is in agreement with an earlier investigation of the complete ONC in which \citet{1998AJ....116.1816H}
report a disc fraction of 50-90\,\% (though relying only on $I_{\mathrm{C}}-K$ colors) and justifies the assumption of a 100\,\% primordial disc
fraction.

The treatment of binaries in our model of the ONC is crucial, since they can have a strong impact on the evolution of cluster dynamics. In the next
section we give a summary of the main properties of the ONC binary population that determine our model setup.

\subsection{Binaries in the ONC}

\label{sec:onc_bin}

In the ONC the binary rate for solar-type stars is $\sim$50\,\%.
From observations alone we have only a very limited knowledge of the distribution of binary periods, eccentricities or mass ratios of the
ONC. However, combining observational data and numerical simulations, the initial properties of the primordial binary population in a stellar aggregate
have been modelled by \citet{1995MNRAS.277.1507K} and \citet{2007A&A...474...77K}. These can be applied to some degree to the ONC.

The investigation of \citet{1995MNRAS.277.1507K} is based on the properties of the Taurus-Auriga binary population and constructs the primordial
population by \emph{inverse dynamical population synthesis} \citep[see][]{1995MNRAS.277.1491K} and \emph{pre-main-sequence eigenevolution}. The
resulting distributions are approximately the log-normal period distribution $f_P(P)$ of \citet{1991A&A...248..485D},
\begin{eqnarray}
f_P(P) \propto \exp{\left[ \frac{\log{P} - \mu_P}{2\sigma_P^2} \right]}\,,  & 
\qquad  P_{\rm{min}} \le P \le P_{\rm{max}},
\end{eqnarray}
with mean $\mu_P$\,$\equiv$\,$\overline{\log{P}}$\,=\,4.8 and standard deviation $\sigma_P$\,$\equiv$\,$\sigma_{\log{P}}$\,=\,2.3, $P$ in days,
a thermally relaxed eccentricity distribution $f_e(e)$,
\begin{eqnarray}
f_e(e)=2e,  &  \qquad  0 \le e < 1, \label{eqn:dist_ecc_therm}
\end{eqnarray}
and a mass ratio distribution $f_q(q)$ obtained by random pairing of stars,
\begin{equation}
f_q(q) \propto q^{\gamma_q}, \label{eqn:dist_mass_ratio}
\end{equation}
where $q$\,=\,$M_2/M_1$, $M_1$ the primary, $M_2$ the secondary mass, and $\gamma_q$\,=\,$\alpha$, $\alpha$ the slope of the mass function of the stellar system.

The log-normal period distribution $f_P(P)$ results in an approximately log-normal semi-major axis distribution $f_a(a)$, the shape of which is slightly
dependent on the distribution over binary mass~$M$ \citep{2007A&A...474...77K}:
\begin{equation}
\overline{\log{a}} = \frac{2}{3} \overline{\log{P}} - \frac{1}{3} \log{\left( \frac{4\pi}{2GM} \right)}, \\
\sigma_{\log{a}} = \frac{2}{3} \sigma_{\log{P}}.
\end{equation}
\citet{2007A&A...474...77K} have analyzed the current binary population of Scorpius OB2, under the reasonable assumption that it is still close to its
primordial state. Accounting for different observational biases by means of comparison with simulated observations of model associations, they recovered
a somewhat different primordial binary population: The semi-major axis distribution of Sco~OB2 is most consistent with a flat distribution in logarithmic
space, and is equivalent to
\begin{eqnarray}
f_a(a) \propto a^{\gamma_a}  &  \qquad  a_{\rm{min}} \le a \le a_{\rm{max}},
\end{eqnarray}
with  $a_{\rm{min}}$\,$\approx$\,5\,{\mbox{R$_{\odot}$}}, $a_{\rm{max}}$\,$\approx$\,5\,$\cdot$\,$10^6$\,{\mbox{R$_{\odot}$}}, and $\gamma_a$\,=\,$-1$, which is also known as \"{O}piks law
\citep{1924TarObs..6...25S}.
The eccentricity distribution could not be well constrained, but the observations are consistent with a thermal distribution, given by
Eq.~(\ref{eqn:dist_ecc_therm}). Unlike \citet{1995MNRAS.277.1507K}, \citet{2007A&A...474...77K} find a power law dependence of the mass ratio
distribution with $\gamma_q$\,$\approx$\,$-0.4$. This is much flatter than Eq.~(\ref{eqn:dist_mass_ratio}) and favors massive companions for massive stars.

The model of \citet{2007A&A...474...77K} seems more applicable to our case, as observational studies of the ONC favor a flat distribution of
the semi-major axes \citep[e.g.][]{1997ApJ...477..705P,2007AJ....134.2272R}.
Theoretical considerations based on three-body encounters give similar results \citep{1997ApJ...485..785V}.

A thermal eccentricity distribution is expected from energy equipartition as a result of multiple soft encounters \citep{1975MNRAS.173..729H} and is
also found from observations, though only for binaries with separations $a$\,$\gtrsim$\,10-50\,AU. Very close systems are subject to circularization due to tidal
effects occurring during stellar evolution \citep{1991A&A...248..485D,1994ARA&A..32..465M}.

The shape of the observed mass ratio distribution is not well constrained by observations.
However, to a good approximation the mass ratio distribution can be described by a power
law as given by Eq.~(\ref{eqn:dist_mass_ratio}) over a wide mass range
\citep[e.g.][]{1990MNRAS.242...79T,2001MNRAS.321..149M,2001AJ....122.1007R,2002A&A...382...92S,2004RMxAC..21..147V}. We thus favor a single mass ratio
distribution for primordial binaries given by Eq.~(\ref{eqn:dist_mass_ratio}) with $\gamma_q$\,=\,$-0.4$ as derived by \citet{2007A&A...474...77K}.

Binaries have strong effects on the overall cluster dynamics mainly through close interactions with single stars or other multiples. In the following we
estimate the typical encounters that could generate high-velocity stars.

\subsection{Three-body encounters in the ONC}

\label{sec:onc_enc}

As is well known, the non-hierarchical motion of three bodies, known as the three-body problem, has no analytical solution, and the chaotic motion of the
members can only be investigated numerically in a statistical manner \citep[see][]{1991ARA&A..29....9V}. In her study of triple systems with negative
total energy, \citet{1986Ap&SS.124..217A} found that about 95\,\% of three-body systems decay after a close triple approach of the components. In most
cases ($\sim$80\,\%), ejection leads to escape, but can also result in the formation of a hierarchical triple system, with one body in an extended
orbit. The lowest mass member has the highest probability of being ejected, about 80\,\%.

In the following we assume that high-velocity stars are typically generated in triple systems with negative total energy. We will
  justify this assumption later in this section. If the motion has not been significantly perturbed since the encounter, one can draw conclusions
about the underlying encounter parameters from the dynamics of the ejected body. In the case of the ONC, it is valid to assume that high-velocity stars
with velocities $\upsilon$\,$\ge$\,3{\mbox{$\sigma_{\rm{3D}}^{\rm{JW}}$}}\,$\approx$\,13\,{\mbox{km\,s$^{-1}$}} are effectively unperturbed before they escape from the cluster (see Appendix
\ref{app:enc_escape}). Then the time to reach the outskirts of the cluster is $t_{\rm{esc}}$\,$\approx$\,$R/\upsilon$\,$\lesssim$\,0.2\,Myr.

For a crude estimate of the compactness of the three-body system (with negative total energy) from which a member is ejected with
$\upsilon$\,$\ge$\,3{\mbox{$\sigma_{\rm{3D}}^{\rm{JW}}$}}, we evaluate the scaling of the median escape speed from \citet{1995A&A...304L...9S},
\begin{equation}
\langle \upsilon_{\rm{esc}} \rangle \approx \frac{1}{2} \left( \frac{|E_0|}{\langle m_{\rm{esc}}\rangle} \right),
\end{equation}
where ${\langle m_{\rm{esc}}\rangle}$ denotes a weighted mean of the escaped particle masses, $|E_0|$\,$\propto$\,$ M^2_{\rm{tot}}/R$ is the total system energy, and
$M_{\rm{tot}}$ and $R$ are the total mass and the scale length of the system.

We assume that the encounter occurred in the dense Trapezium Cluster where it is most probable \citep[see Fig.~4
of][]{2006A&A...454..811P}. Due to mass segregation of the cluster, the mass of the most massive component of the three-body system is likely to be
several times the mean stellar mass in the ONC; we adopt $M_{\rm{tot}}$\,=\,4\,{\mbox{M$_{\odot}$}} for the system mass. The mass of the ejected body is assumed to be
half the mean stellar mass, ${\langle m_{\rm{esc}}\rangle}$\,=\,0.25\,{\mbox{M$_{\odot}$}}. With these assumptions we obtain $R$\,$\lesssim$\,100\,AU for the scale length of the system.
However, the minimum approach which eventually leads to an ejection will be much closer. Since the disc radius of a low-mass star is about
$r_d$\,$\approx$\,100\,AU, the ejected component can lose more than 90\,\% of the disc mass \citep[see Table 3 of][]{2006ApJ...642.1140O}.

Initially we assumed that triple systems that generate high-velocity stars have negative total energy. In Appendix~\ref{app:max_vel_triple_escaper}
  we show that this is a valid assumption if the ejected high-velocity stars do not exceed velocities of a few tens of {\mbox{km\,s$^{-1}$}}. In fact, all of the
  high-velocity stars from numerical simulations and observational data do not show higher velocities. A detailed analysis of our numerical
  simulations shows that in all cases the encounter-generated high-velocity stars are the lowest mass component ejected from a three-body
  system with a massive tight binary ($M_\mathrm{bin}$\,$\gtrsim$\,20\,{\mbox{M$_{\odot}$}}, $a_\mathrm{bin}$\,$\lesssim$\,50\,AU), leaving on a nearly parabolic
  orbit relative to one of the binary components. We thus conclude that the triple systems which generated the observed low-mass high-velocity stars must
  have been bound as well.

How many such encounters do we expect in the ONC? For simplicity we assume the three-body interaction to result from a single-binary encounter, where the
binary has a semi-major axis $a$\,$\lesssim$\,100\,AU. As a volume relevant for close encounters we consider the Trapezium Cluster. Referring to
  Eq. (\ref{eqn:tcoll}) the time scale for a three-body encounter at 100\,AU is $t_{\rm{enc, 100\,AU}}$\,$\approx$\,10\,Myr. Since the time the star
remains in the cluster volume is $t_{\rm{esc}}$\,$\lesssim$\,0.2\,Myr, the probability of the detection of a high-velocity star is
$P_{\rm{hvs}}$\,$\approx$\,$t_{\rm{esc}} / t_{\rm{enc,100\,AU}}$\,$\lesssim$\,0.02. Knowing the number of stars located in the Trapezium Cluster,
$N_{\rm{TC}}$, we expect $N_{\rm{hvs}}$\,$\approx$\,$P_{\rm{hvs}} N_{\rm{TC}}$\,$\lesssim$\,15 high-velocity stars in the ONC.

In summary, high-velocity stars with velocities $\upsilon$\,$\ge$\,3{\mbox{$\sigma_{\rm{3D}}^{\rm{JW}}$}} (or velocity components $\upsilon_{x,y,z}$\,$\ge$\,3{\mbox{$\sigma_{\rm{1D}}^{\rm{JW}}$}}) have most
probably experienced exactly one close ($\lesssim$\,100\,AU), disruptive encounter in the cluster centre and leave the cluster on a radial
trajectory, i.e. with the velocity vector pointing away from the cluster centre, preserving the dynamical information of the encounter. We
expect about one dozen such stars in the ONC.

In the following we will compare the observationoal data and theoretical estimates with numerical simulations of a dynamical model of the ONC.

%

\section{Numerical simulations}

\label{sec:sim}

The basic dynamical model of the ONC used here is described in \citet{2006ApJ...642.1140O}, with several extensions discussed in
\citet{2007A&A...475..875P}. We summarize the main properties of the initial stellar distribution: the simulations start in virial equilibrium,
  $Q$\,=\,0.5, with a radial density profile, $\rho \propto r^{-2}$, a central density $\sim$4$\cdot10^5$\,pc$^{-3}$, and a Maxwellian velocity distribution.
Here we include additional effects like a varying background potential and a population of primordial binaries. We have performed 20 runs with different
random configurations of positions and velocities from the given distributions to establish a statistically robust database. Unless explicitly declared
otherwise the presented results refer to the whole set of runs. The reason to extend our basic model of the ONC is that we want to compare the velocity
distribution to the observational data. Both the background potential and primordial binaries have a large impact on the resulting velocity
distribution. However, for comparison and analysis of the contribution of the background potential and primordial binaries, we also performed
simulations of a single particle model, a single particle model with a background potential, and a cluster with a primordial binary population but
without a background potential.

A background potential increases the cluster virial mass and thus results in a higher velocity dispersion (cf. Eq. (\ref{eqn:vir})). In our simulations
the background potential is represented by a Plummer sphere,
\begin{equation}
\rho(r)=\frac{3M}{4\pi a^3}\frac{1}{(1+\frac{r^2}{a^2})^{5/2}},
\end{equation}
with mass $M$ and characteristic length scale $a$. The Plummer sphere is set up with 6000\,{\mbox{M$_{\odot}$}} initially and a mass loss rate
$\dot{M}$\,=\,4000\,{\mbox{M$_{\odot}$}}\,Myr$^{-1}$. Due to the continuous mass loss, after 1\,Myr, 2000\,{\mbox{M$_{\odot}$}} of gas is left. The exact time-scale and
time dependency of the mass loss is not crucial, as long as the gas expulsion time scale is of the order of the dynamical time scale of the system,
$t_{\rm{exp}}$\,$\approx$\,$t_{\rm{dyn}}$, which appears to be the case for the ONC (cf. \S\ref{sec:onc_dyn}). The length scale is set to $a$\,=\,0.6\,pc to match
the observed velocity dispersion at a simulation time of $t_{\rm{sim}}$\,=\,1\,Myr.

The effect of primordial binaries is more complicated. The interactions of binaries and single
stars or other binaries have the potential to change the velocity distribution much more than a single star model. In particular, three
body encounters between singles and binaries usually lead to the expulsion of the lowest mass member from the small $N$-body system with a high
velocity. This mechanism is especially important for the high-velocity fraction of particles of interest in this study (see~\S\ref{sec:onc_enc}).

We have set up a series of simulations with the models of \citet{1995MNRAS.277.1507K} and \citet{2007A&A...474...77K} and found that results do not
depend on the choice of one particular model. The initial binary frequency was chosen to be 75\,\%. Although the observed binary frequency in the ONC
is $\sim$50\,\%, it is necessary to start with a higher binary rate as about one third of the binaries are distroyed within the first 1\,Myr due to
dynamical evolution.

\begin{figure}[t]
   \centering
   \includegraphics[height=1.0\linewidth,angle=-90]{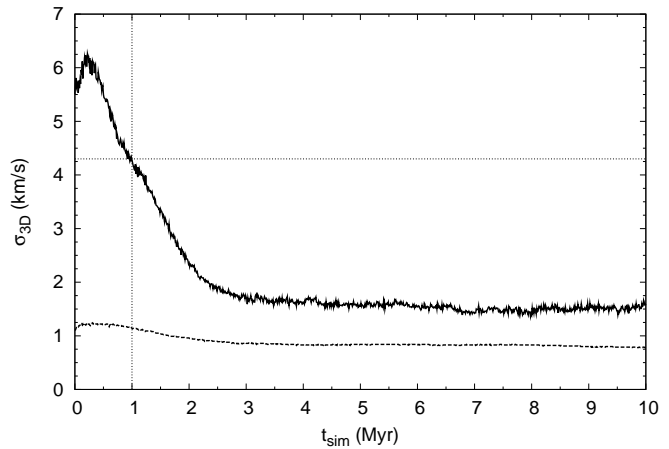}
   \caption{Three-dimensional velocity dispersion as a function of time for the previous (dashed) and new (solid) numerical models of the ONC. The
     dotted horizontal line marks the three-dimensional velocity dispersion from \citet{1988AJ.....95.1755J}, {\mbox{$\sigma_{\rm{3D}}^{\rm{JW}}$}}\,=\,4.3\,{\mbox{km\,s$^{-1}$}}, the vertical dotted
     line marks the mean age of the ONC, $t_{\rm{ONC}}$\,=\,1\,Myr.}
   \label{fig:vel_disp}
\end{figure}

In terms of global cluster dynamics, the new model with a background potential and primordial binaries provides a much better fit to the ONC data than
the previous single star model without background potential. We illustrate this by means of the time-evolution of the three-dimensional velocity
dispersion of both models in Fig.~\ref{fig:vel_disp}. In the previous model there was not enough mass confined in the cluster to reproduce the observed
velocity dispersion of the ONC, {\mbox{$\sigma_{\rm{3D}}^{\rm{JW}}$}}\,=\,4.3\,{\mbox{km\,s$^{-1}$}}, at $\sim$1\,Myr. The new model gives a much better result. The continuous, steep falloff after
$\sim$0.2\,Myr is due to the response of the stellar system to the gas expulsion. In the following, we will discuss results of numerical simulations based
on the new dynamical model of the ONC.

As demonstrated in \citet{2006ApJ...642.1140O}, \citet{2006A&A...454..811P}, and \citet{2007A&A...462..193P}, stellar encounters in dense clusters can
lead to significant transport of mass and angular momentum in protoplanetary discs. In the present investigation we have used Eq.~(1) from
\citet{2006A&A...454..811P} to keep track of the disc-mass loss of each star due to encounters. Our estimate of the
  accumulated disc-mass loss is an upper limit because the underlying formula is only valid for co-planar, prograde encounters, which are
  the most perturbing. A simplified prescription assigns stars into one of two
distinct groups: if the relative disc-mass loss exceeds 90\,\% of the initial disc mass, stars are marked as ``discless''; otherwise they are termed
``star-disc systems''. This approach is justified by the interplay of three effects: (i) A disc-mass loss of this order lowers the density in the disc
significantly, in particular in the outer parts; the disc size decreases. (ii) The accompanying angular momentum loss enhances accretion of the extant
material onto the star \citep{2006ApJ...652L.129P}. This leads temporarily to an increase of the infrared excess but soon fades after a short intense
accretion phase (Pfalzner (2008), subm. to A\&A). (iii) The loose distribution of circumstellar matter lowers the shielding of the disc midplane against
photoevaporation. The interplay of these effects leads to a fast dispersal of the disc material. From the observational point of view, the corresponding
star would show pure photospheric emission on the order of some $10^3$\,yr after the encounter.

\begin{figure}[t]
   \centering
   \includegraphics[height=1.0\linewidth,angle=-90]{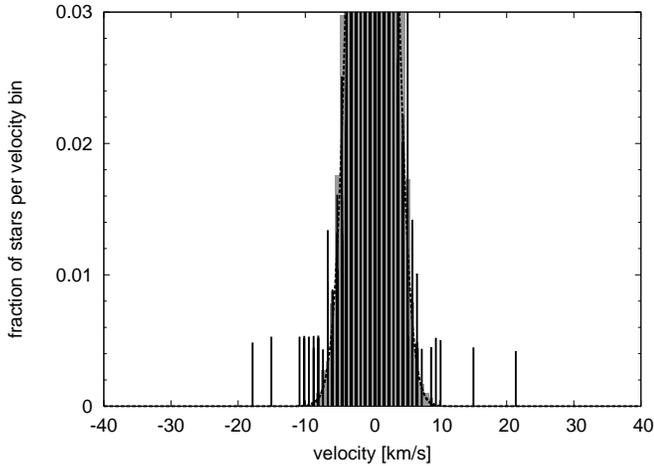}
   \caption{Velocity distribution of cluster stars from simulations. The construction of the distribution is described in the text. The bin
     width is 0.7\,{\mbox{km\,s$^{-1}$}}. The stars have been divided into two groups according to their extant disc mass (see text): boxes represent star-disc systems,
     impulses represent discless stars. For comparison a Gaussian with dispersion $\sigma$\,=\,{\mbox{$\sigma_{\rm{1D}}^{\rm{JW}}$}}\,=\,2.5\,{\mbox{km\,s$^{-1}$}} is superposed (dashed line).}
   \label{fig:gauss_num}
\end{figure}

In Fig.~\ref{fig:gauss_num} the velocity distribution of cluster stars is shown after 1\,Myr of evolution. This is done separately for discless stars and star-disc
systems. For comparison with the observational data presented in Fig.~\ref{fig:obs_gauss}c, separate velocity distributions of two spacial directions (here $x$
and $y$) have been added to mimic the distribution of proper motion data. Unless in wide systems, primaries and secondaries could not have been
resolved by \citet{1988AJ.....95.1755J} who worked with seeing-limited images. So the presence of binaries in the numerical simulations requires a special
treatment of velocities. Accounting for unequal mass components and nebulosity, we adopt a minimum separation of 1000\,AU \citep[corresponding to
$\sim$2.5$''$; see][and references therein]{2008arXiv0801.4085M} for the visual resolution of a binary system. For closer systems, only the primary component
is taken into account and the center-of-mass velocity of the system is assigned. For wider systems, both components are treated as single stars. This
prescription is simple and rough but appropriate to avoid the inclusion of large velocity components in tight binaries.

The velocity distribution shows common features with Fig.~\ref{fig:obs_gauss}c, which was obtained from observational data: the bulk of the stars forms a
relaxed system which manifests in the approximate Gaussian velocity distribution with a characteristic velocity dispersion $\sigma$\,=\,{\mbox{$\sigma_{\rm{1D}}^{\rm{JW}}$}}. Moreover, a
small fraction of stars exists with much higher velocities $\upsilon_{x,y}$\,$\ge$\,3{\mbox{$\sigma_{\rm{1D}}^{\rm{JW}}$}}, the previously described ``high-velocity stars''.

The velocity distributions of stars that have lost their disc due to close encounters and those that have retained their disc differ. The discless
population consists of a larger fraction of high-velocity stars, while the width of the Gaussian part is similar. This feature is in agreement with
expectations: High-velocity stars are usually the lowest mass members of temporary few-body systems which are expelled after a close encounter (see
\S\ref{sec:onc_enc}). The close passage and large mass of the perturber results in a significant removal of disc material
\citep{2006A&A...454..811P}.

\begin{figure}[t]
   \centering
   \includegraphics[height=1.0\linewidth,angle=-90]{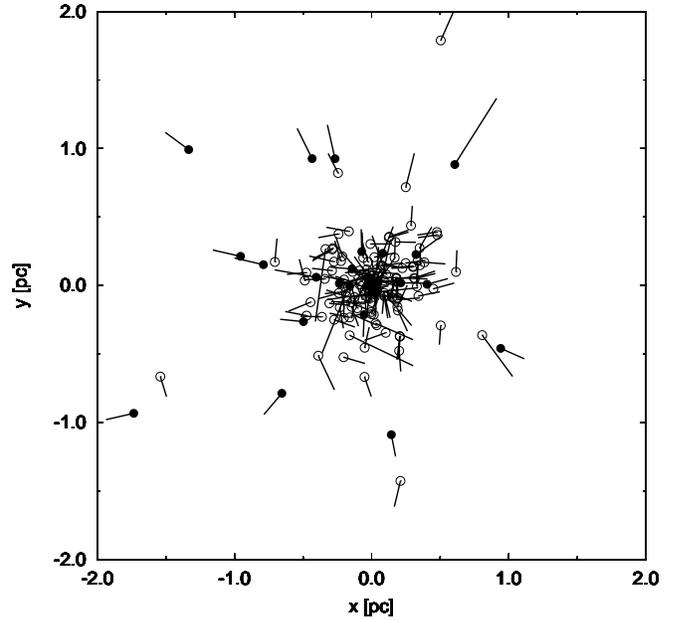}
   \caption{Positions and velocities of high-velocity star-disc systems (open circles) and discless stars (filled circles) from cluster simulations.
   The frame is centered on the cluster centre. The lines indicate the distance (in pc) a star would have moved in a period of 2\,$\cdot$\,$10^4$\,yr.}
   \label{fig:num_pos_vel}
\end{figure}

In analogy to Fig.~\ref{fig:obs_pos_vel}, positions and velocities of high-velocity stars from the numerical simulations are displayed in
Fig.~\ref{fig:num_pos_vel}. Here the great advantage of numerical simulations becomes apparent: several runs of the same model can improve
statistics far enough to produce prominent features where only weak signatures in observational data are found. Our dynamical model of the ONC reproduces
the observed features: As expected, most stars are concentrated in the inner tenth parsecs around the cluster centre, while several stars are located in the
outer cluster parts, moving in radial directions from the cluster centre. However, we find the same two unexpected features as in the
observations, namely that (i) the region between the cluster centre and the outer cluster parts is underpopulated by high-velocity stars, and (ii) the
velocity vectors of a fraction of stars in the outer cluster parts do not point away from the cluster centre. We will refer to high-velocity
  stars that leave the cluster on a track in radial direction from the cluster centre, i.e. with the velocity vector pointing away from the cluster core
  with radius $R_\mathrm{in}$\,$\approx$\,0.1\,pc, as ``radial escapers'', while high-velocity stars that do not match this condition will be called
  ``orbital escapers''. In the following we explain the choice of the terminology and discuss the two classes of high-velocity stars.

\begin{figure}[t]
   \centering
   \includegraphics[height=0.8\linewidth,angle=-90]{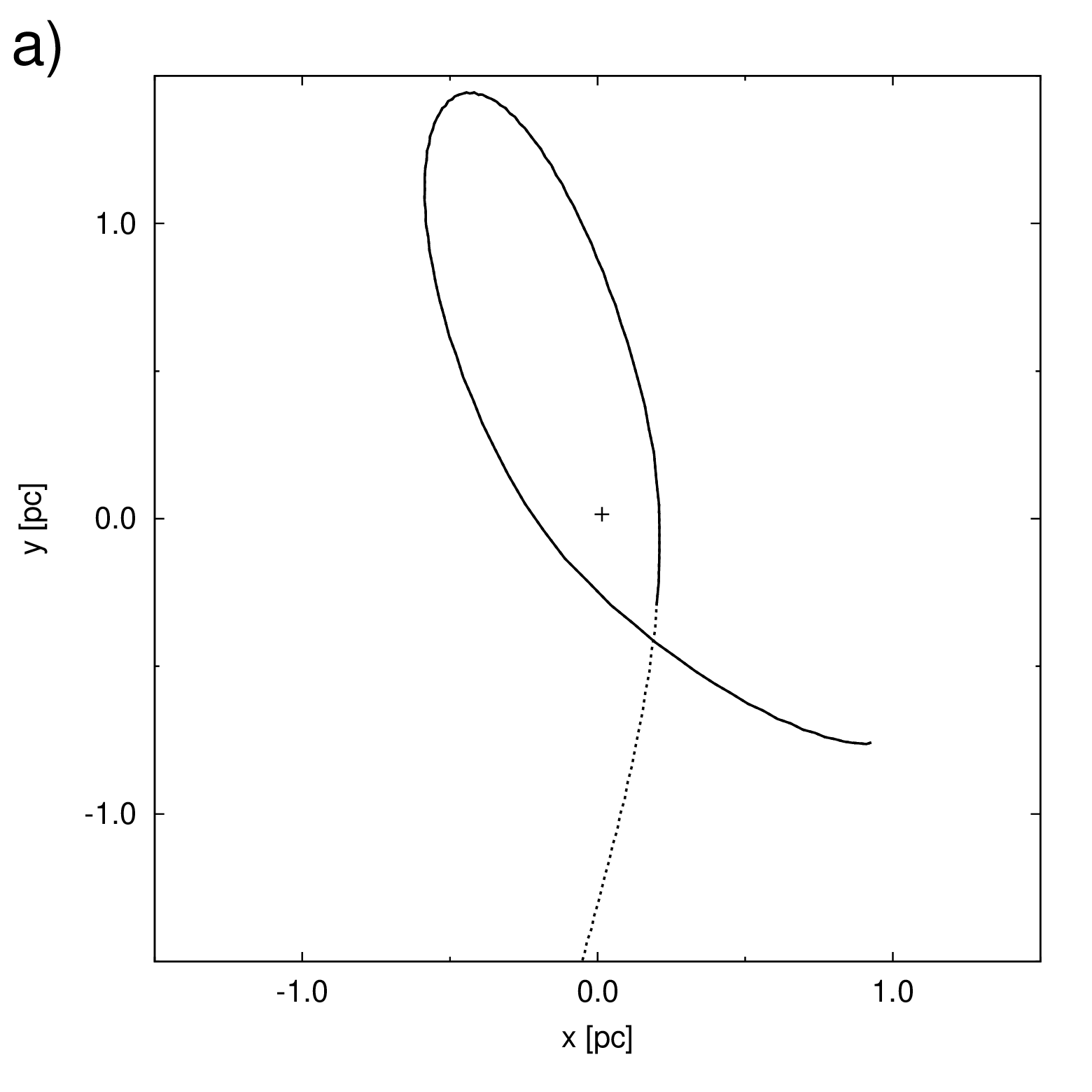}
   \includegraphics[height=0.8\linewidth,angle=-90]{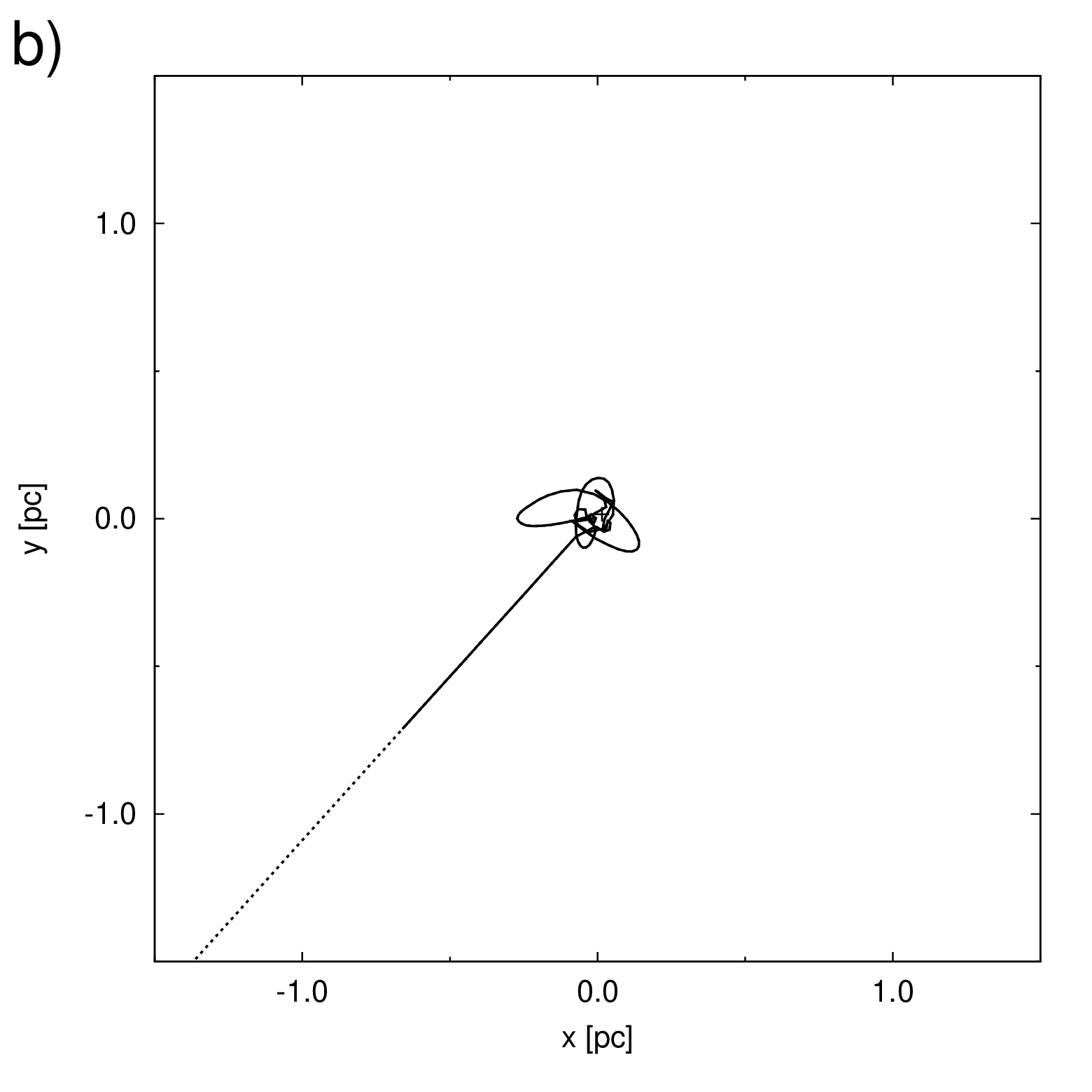}
   \caption{Tracks of two high-velocity stars from simulations. The cross marks the cluster centre. The solid line marks the trajectory up to 1\,Myr,
     the dotted line represents later times. {\textbf{a)}} Star on a wide orbit around the cluster centre, escaping from the cluster after a close
     passage of the cluster centre. {\textbf{b)}} Star escaping after multiple passages of the cluster centre and a final close encounter.}
   \label{fig:tracks_example}
\end{figure}

In contrast to observations we can trace the history of these stars in the simulation; this gives us the opportunity to investigate the reason for this strange
configuration. As we will see, (i) is a consequence of (ii), thus we will discuss that first. In Fig.~\ref{fig:tracks_example} two exemplary tracks of stars in
the outer cluster parts are displayed, those of a radial (Fig.~\ref{fig:tracks_example}b) and an orbital escaper (Fig.~\ref{fig:tracks_example}a). It is
evident that the phase spaces of the two stars are completely different. The radial escaper was originally located close to the cluster centre,
experienced several encounters in multiple passages of the cluster centre and was finally expelled in a close three body encounter. After the
  ejection, it is moving on a track in a radial direction from the cluster centre, i.e. with the velocity vector pointing away from the cluster core. The orbital
escaper stems from the outer cluster parts, passed on a non-closed orbit at a minimum distance of some tenths of a parsec around the cluster centre
(without significant encounters) and was accelerated sufficiently by the central mass to leave the cluster on a hyperbolic orbit. Most of the
  time its velocity vector is \emph{not} pointing away from the cluster core. This is most evident at large distances from the cluster
  centre. Only at the two short periods of cluster centre passage, i.e. when the radius vector is approximately normal to the velocity vector, the
  determination of the direction of motion is eventually not sufficient to discriminate between a radial and orbital escaper. The fact that the orbital escaper is
  leaving the cluster although it was initially energetically (but only weakly) bound to the cluster is due to a varying cluster potential on a time scale shorter than its
  revolution around the cluster centre.\footnote{The time scale of the potential variation is related to the crossing time of the cluster,
    $t_\mathrm{cross} \approx 0.5$\,Myr, while the time scale of revolution is approximately the cluster age, $t_\mathrm{ONC} \approx 1$\,Myr, which is
    about twice as large.} The main sources of the potential variation are mass segregation and the evaporation of (preferentially low-mass) cluster
  stars. Consequently the orbital escaper is accelerated more strongly after the second passage of the cluster centre and is less decelerated in the outer
  parts due to the lower total cluster mass and more extended cluster potential.

The existence of two different classes of tracks explains why the region between the cluster centre and the outer cluster parts is devoid of high-velocity
stars. Radial escapers leave the cluster on a very short time scale, while orbital escapers reside for much longer in the vicinity of the cluster centre due to
their wide non-closed orbits. Hence the disjoint spacial groupings of stars after 1\,Myr of cluster evolution are the consequence of disjoint sets of
initial phase space volumes.

We find from our simulations that the distinct dynamics of the two exemplary high-velocity stars characterize \emph{in general} the dynamics of radial and
orbital escapers: they belong to dynamically distinct groups. Moreover, the two groups as well can be separated due to their
disc properties: a large fraction of radial escapers is discless, while most orbital escapers are star-disc systems. This \emph{morphological}
distinction is a consequence of the dynamical bisection. The morphological bisection translates observationally into a photometrical bisection, i.e. the
stars would be devided into two groups according to the presence of excess emission. Such a trend, though only weak, is also present in the
observational data.

In Fig.~\ref{fig:obs_pos_vel}b, the three isolated stars with excess emission are identified as orbital escapers. The situation is more diffcult
  for the two stars with excess emission immediatly below the cluster centre. As mentioned above, stars close to the cluster centre cannot be uniquely identified as radial
  or orbital escapers from the direction of motion alone. Moreover, we have only two-dimensional spacial and velocity information, so the true distances
  to the cluster centre and velocities are not known. If the projected distance and true distance to the cluster centre differ only slightly for both
  stars, then we would classify the more distant ($\sim$0.25\,pc) as a probable orbital escaper. The reason is that due to its proximity to the
  cluster centre and relatively low velocity this star will be accelerated and deflected by the central cluster mass and pass on a curved
  trajectory. When passing the cluster outskirts, its direction of motion would not point away from the cluster core and thus it would be
  identified as an orbital escaper according to our classification scheme. Of course, we cannot exclude the possibility that this star was ejected in a
  three-body encounter, although it could not be classified as a radial escaper due to its predicted trajectory. For the other star a
  classification as a radial escaper seems more appropriate. If, on the contrary, the true distance is much larger than the projected distance for both stars,
  they would be classified as orbital escapers. However, we can only speculate about the dynamical origin of the two stars. The special case of the two
  close-by stars with excess emission in the lower left will be discussed below. The only star without
excess emission (JW~45) seems to have been expelled very close ($<$\,0.13\,pc) to the cluster centre and is thus classified as a radial
  escaper. If we thus interpret, conversely to the previous arguments, the lack of excess emission as an indicator of the absence of a disc, then this
star provides evidence for encounter-triggered disc destruction. The signature of disc material of the other stars is - as far as a classification is
possible - in accordance with our numerical results and thus supports this view.

The two close-by stars at approximately $(-1.5\,{\rm{pc}}, -1.0\,{\rm{pc}})$ in Fig.~\ref{fig:obs_pos_vel}, JW~3 and JW~4, seem to form
a binary. Though their separation of about $10^4$\,AU is large, the remarkably similar proper motions and radial velocity
\citep[$\upsilon_r^{\mathrm{JW\,3}}$\,=\,29.1\,{\mbox{km\,s$^{-1}$}}, $\upsilon_r^{\mathrm{JW\,4}}$\,=\,31.6\,{\mbox{km\,s$^{-1}$}}:][]{1999AJ....117.2941S}, age, and infrared excess
strongly support the assumption of a physical pair - at least in the past. If this pair was expelled as a binary from a four-body encounter, than this
must have occurred less than 0.1\,Myr ago (accounting for the actual distance from the cluster centre, the velocity and the deceleration by the interior
cluster mass). The difference in proper motion of the two stars corresponds to a distance of $\sim(4\pm2)$\,$\cdot$\,$10^4$\,AU, in good agreement with
the observed projected separation. Due to the direction of motion, the two stars are classified as radial escapers. The expulsion of binaries from
close four-body encounters in our simulations, though a rare event (9 events from our 20 runs), usually does not lead to a significant disc-mass loss of
the individual stars. This again is in good agreement with the observational data. Thus the excess emission of the binary radial escaper does not
contradict the correlation of our dynamical and the photometric classification, which in the case of single stars shows that in most cases radial
escapers are discless and orbital escapers are found preferentially among star-disc systems.

\begin{figure}[t]
   \centering
   \includegraphics[height=1.0\linewidth,angle=-90]{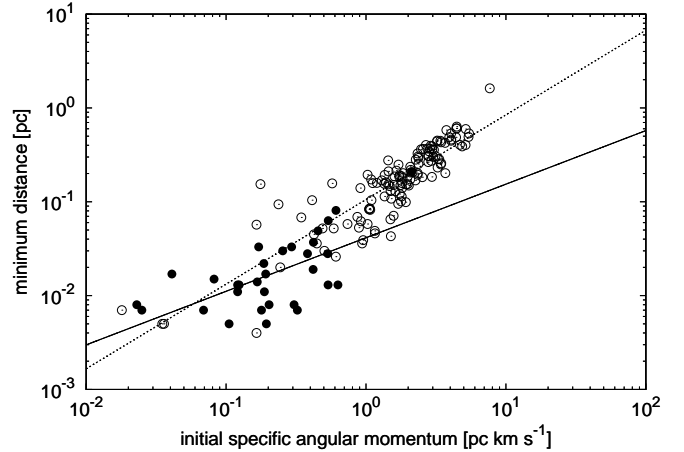}
   \caption{Minimum distance of the high-velocity stars to the cluster centre, $d_{\mathrm{min}}$, as a function of the initial specific angular
     momentum, $l_0$ (relative to the cluster centre). Star-disc systems are marked by open, discless stars by filled circles. Linear best-fits of both
     populations, star-disc systems (dotted line) and discless stars (solid line), have been included (quantitative results are discussed in the text).}
   \label{fig:track_dmin_l0}
\end{figure}

The general difference of the orbits of discless stars and star-disc systems is represented in Fig.~\ref{fig:track_dmin_l0}. The minimum distance of
star-disc systems to the cluster centre $d_{\rm{min}}$ is clearly a linear function of the initial specific angular momentum $l_0$ (relative to the
cluster centre). From a linear least squares fit we find a slope $\alpha$ with a small standard error $\Delta\alpha$,
  $\alpha$\,=\,(0.902\,$\pm$\,0.044)\,s\,km$^{-1}$ ($\Delta\alpha/\alpha$\,=\,0.048). This means that $l_0$ is conserved - a consequence of the
wide orbit around the cluster centre without strong, abrupt perturbations from single stars. On the contrary, discless stars show a much wider, not
  clearly correlated distribution of $d_{\rm{min}}$ due to angular momentum exchange in close encounters in the cluster centre. Here the slope
  $\beta$ of the linear best-fit has a large standard error $\Delta\beta$, $\beta$\,=\,(0.571\,$\pm$\,0.127)\,s\,km$^{-1}$
  ($\Delta\beta/\beta$\,=\,0.222). Those star-disc systems with lower $l_0$ and $d_{\rm{min}}$, populating the discless regime, are components of
binaries.

\begin{figure}[t]
   \centering
   \includegraphics[height=1.0\linewidth,angle=-90]{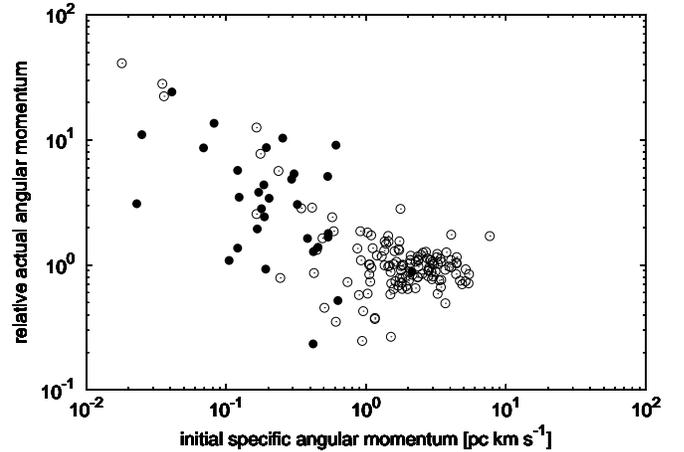}
   \caption{Relative actual angular momentum of the high-velocity stars, $l_{\mathrm{now}}/l_0$, as a function of the initial specific angular momentum
     $l_0$ (relative to the cluster centre). Star-disc systems are marked by open, discless stars by filled circles.}
   \label{fig:track_lnowl0_l0}
\end{figure}

The distinct dynamics are even more evident from Fig.~\ref{fig:track_lnowl0_l0}: here we show the ratio of the actual (at 1\,Myr) and initial specific angular momentum
$l_{\rm{now}}/l_0$ as a function of the initial specific angular momentum $l_0$. Star-disc systems are concentrated nearly symmetrically around $l_{\rm{now}}/l_0$\,=\,1, while for most discless stars $l_{\rm{now}}/l_0$\,$>$\,1, and even up to several tens. The increase of angular momentum of the discless stars \emph{in
relation to the cluster centre} can be explained as follows: Single stars gain a large amount of angular momentum in a close triple encounter
\citep{2005MNRAS.364...91V}, and leave on straight radial tracks after breakup. Since these
encounters occur preferentially close to the cluster centre, on average the angular momentum relative to the cluster centre is also highly
increased. The raise of $l_{\rm{now}}/l_0$ with lower $l_0$ is given by the fact that close encounters are more probable for stars with lower $l_0$,
which in turn lead to a higher gain in angular momentum.

The additional components in our numerical model of the ONC - a background potential and primordial binaries - have different effects on the sample of
high-velocity stars. By comparing with results of our simulations without either one or both additional components, we find that (i) the inclusion of
primordial binaries has the effect of increasing the number of discless stars (due to a larger number of encounters) and to increase the maximum velocity
of the high-velocity stars (due to a higher probability of closer encounters), (ii) a background potential reduces the number of discless stars
and reduces the maximum velocity of the high-velocity stars, because the higher velocity dispersion of the stars reduces the probability of
close encounters. The combination of both as in the present model does not cancel out the positive effect of the binaries and thus results in a
\emph{higher} number of discless stars and a higher maximum velocity of the high-velocity stars than would result from a single star model.

%

\section{Summary and discussion}

\label{sec:discussion}

Combining observational data and numerical simulations, we have shown that even after 1\,Myr of dynamical evolution a cluster of the size and density of
the ONC is still dynamically active. Young stars, preferentially of low-mass, are expelled in close $N$-body encounters, losing a large fraction of their
circumstellar matter.

The encounters are in most cases interactions with massive stars in the cluster centre. This finding highlights the significant
effect that encounters in (massive) young stellar clusters can have on the evolution of protoplanetary discs. This is even more evident if one addresses
not only the effect on the mass, but the even stronger effect on the angular momentum of the disc as shown in previous investigations
\citep{2006ApJ...652L.129P,2007A&A...462..193P}.

Using the observational data from \citet{1988AJ.....95.1755J} we show (Fig.~\ref{fig:obs_gauss}c) that in the ONC there is a small population of stars with
proper motions $\mu_{\rm{x,y}}$ larger than three times the one-dimensional velocity dispersion of the cluster, $\mu_{\rm{x,y}}$\,$\gtrsim$\,7.5\,{\mbox{km\,s$^{-1}$}}.

From numerical simulations we find that the so-called ``high-velocity stars'' form two dynamically disjoint groups. One group is composed of stars expelled in a close encounter, moving
on radial tracks directly outward from the cluster centre, termed ``radial escapers''. The other contains unperturbed stars running on wide, non-closed orbits around the cluster
centre (on average with lower velocities), termed ``orbital escapers''. The dynamical state of the stars has its origin in their initial
phase space location: radial escapers are initially located close to the cluster centre ($r$\,$\lesssim$\,0.3\,pc) and characterized by a low
angular momentum (relative to the cluster centre). In contrast, orbital escapers were formed in the outer cluster with large angular
momentum relative to the cluster centre.

The different dynamics of the high-velocity stars implies a signature in their circumstellar characteristics: stars being subject to close encounters are
expected to lose their disc material faster and to a higher degree \citep{2006A&A...454..811P}. Using our prescription of disc-mass loss in stellar
encounters, indeed we find a clear correlation from numerical simulations: radial escapers, initially located close to the cluster centre and
later expelled in a close encounter, lose more than 90\,\% of their disc material in 1\,Myr of dynamical evolution, while orbital escapers,
initially distant stars, moving on wide orbits around the cluster centre, do not experience strong perturbations and keep most of their disc material.

We compare our numerical results with infrared observations from \citet{1997AJ....113.1733H}, \citet{2002ApJ...573..366M}, and
\citet{2004AJ....128.1254L}, tracing the (inner) circumstellar material. Though strict conclusions are not possible due to the small observational
sample and some stars that cannot be classified according to our scheme, we observe the same trend: stars that have been classified as
  orbital escapers do show near-infrared excess emission, indicative of circumstellar matter, while those classified as radial escapers show pure photospheric colors,
  lacking evidence of (inner) circumstellar discs.

This view is strongly supported by the work of \citet{2004ApJ...607L..47T} on the dynamics of {\mbox{$\theta^1{\rm{C}}\:{\rm{Ori}}$}}. The most massive star in the ONC, {\mbox{$\theta^1{\rm{C}}\:{\rm{Ori}}$}}, has a proper motion several times
greater than the dispersion of bright ONC stars and much larger than the velocity expected if it were in equipartition with the other cluster stars
\citep{1988AJ.....95.1744V}. \citet{2004ApJ...607L..47T} showed that the direction of {\mbox{$\theta^1{\rm{C}}\:{\rm{Ori}}$}}'s motion is consistent with being exactly opposite to that of
a B-type star embedded in the background molecular cloud, the so-called BN object, and concludes that it is most probably a runaway star originating from
the Trapezium ejected about 4000 years ago after a close encounter with {\mbox{$\theta^1{\rm{C}}\:{\rm{Ori}}$}}. This supports the idea that stellar encounters may be likely events in
such dense regions as the ONC, or at least in their cores. The evidence that the most massive stellar object was involved in a close encounter with
another massive star is in best accordance with the work of \citet{2007ApJ...661L.183M} and \citet{2007A&A...475..875P}.

Considering the dynamical age of the ONC of several crossing times, encounters must have had much stronger impacts on a stellar disc at an early age of
the cluster, when densities were much higher but massive stars already had formed. Indeed, our simulations confirm this expectation, giving rise to an
era of strong and frequent interactions among star-disc systems at the onset of massive star formation.

We expect that in even denser clusters such as the Arches cluster, high-velocity stars should be even more frequent than in the ONC. In such
systems the current spatial and velocity distribution of the high-velocity stars should give strong indications of how the cluster developed in former times.

%



%

\begin{acknowledgements}
We thank the anonymous referee for careful reading and very useful comments and suggestions which improved this work. We are grateful to B. Jones for
providing us the proper motion data of the ONC. We also thank R. Spurzem for providing the {\textsc{\mbox{nbody6\raise.2ex\hbox{\tiny{++}}}}} code for the cluster simulations. Simulations were
partly performed at the John von Neumann Institute for Computing, Research Centre J\"{u}lich, Project HKU14. This research has made use of the SIMBAD
database and the VizieR catalogue access tool, operated at CDS, Strasbourg, France.
\end{acknowledgements}

%

\bibliographystyle{aa}
\bibliography{9804ms}

\begin{thebibliography}{71}
\expandafter\ifx\csname natexlab\endcsname\relax\def\natexlab#1{#1}\fi

\bibitem[{{Adams} {et~al.}(2004){Adams}, {Hollenbach}, {Laughlin}, \&
  {Gorti}}]{2004ApJ...611..360A}
{Adams}, F.~C., {Hollenbach}, D., {Laughlin}, G., \& {Gorti}, U. 2004, \apj,
  611, 360

\bibitem[{{Adams} {et~al.}(2006){Adams}, {Proszkow}, {Fatuzzo}, \&
  {Myers}}]{2006ApJ...641..504A}
{Adams}, F.~C., {Proszkow}, E.~M., {Fatuzzo}, M., \& {Myers}, P.~C. 2006, \apj,
  641, 504

\bibitem[{{Anosova}(1986)}]{1986Ap&SS.124..217A}
{Anosova}, J.~P. 1986, \apss, 124, 217

\bibitem[{{Balog} {et~al.}(2007){Balog}, {Muzerolle}, {Rieke}, {Su}, {Young},
  \& {Megeath}}]{2007ApJ...660.1532B}
{Balog}, Z., {Muzerolle}, J., {Rieke}, G.~H., {et~al.} 2007, \apj, 660, 1532

\bibitem[{{Beust} {et~al.}(2005){Beust}, {Reche}, \&
  {Augereau}}]{2005prpl.conf.8092B}
{Beust}, H., {Reche}, R., \& {Augereau}, J.-C. 2005, in Protostars and Planets
  V, 8092--+

\bibitem[{{Binney} \& {Tremaine}(1987)}]{1987gady.book.....B}
{Binney}, J. \& {Tremaine}, S. 1987, {Galactic dynamics} (Princeton, NJ,
  Princeton University Press, 1987, 747 p.)

\bibitem[{{Cabrit} {et~al.}(2006){Cabrit}, {Pety}, {Pesenti}, \&
  {Dougados}}]{2006A&A...452..897C}
{Cabrit}, S., {Pety}, J., {Pesenti}, N., \& {Dougados}, C. 2006, \aap, 452, 897

\bibitem[{{Clarke} \& {Pringle}(1991)}]{1991MNRAS.249..584C}
{Clarke}, C.~J. \& {Pringle}, J.~E. 1991, \mnras, 249, 584

\bibitem[{{Currie} {et~al.}(2008){Currie}, {Kenyon}, {Balog}, {Rieke}, {Bragg},
  \& {Bromley}}]{2008ApJ...672..558C}
{Currie}, T., {Kenyon}, S.~J., {Balog}, Z., {et~al.} 2008, \apj, 672, 558

\bibitem[{{Duquennoy} \& {Mayor}(1991)}]{1991A&A...248..485D}
{Duquennoy}, A. \& {Mayor}, M. 1991, \aap, 248, 485

\bibitem[{{Furesz} {et~al.}(2007){Furesz}, {Hartmann}, {Megeath},
  {Szentgyorgyi}, \& {Hamden}}]{2007arXiv0711.0391F}
{Furesz}, G., {Hartmann}, L.~W., {Megeath}, S.~T., {Szentgyorgyi}, A.~H., \&
  {Hamden}, E.~T. 2007, ArXiv e-prints, 711

\bibitem[{{Haisch} {et~al.}(2000){Haisch}, {Lada}, \&
  {Lada}}]{2000AJ....120.1396H}
{Haisch}, K.~E., {Lada}, E.~A., \& {Lada}, C.~J. 2000, \aj, 120, 1396

\bibitem[{{Haisch} {et~al.}(2005){Haisch}, {Jayawardhana}, \&
  {Alves}}]{2005ApJ...627L..57H}
{Haisch}, Jr., K.~E., {Jayawardhana}, R., \& {Alves}, J. 2005, \apjl, 627, L57

\bibitem[{{Haisch} {et~al.}(2001){Haisch}, {Lada}, \&
  {Lada}}]{2001ApJ...553L.153H}
{Haisch}, Jr., K.~E., {Lada}, E.~A., \& {Lada}, C.~J. 2001, \apjl, 553, L153

\bibitem[{{Heggie}(1975)}]{1975MNRAS.173..729H}
{Heggie}, D.~C. 1975, \mnras, 173, 729

\bibitem[{{Heller}(1995)}]{1995ApJ...455..252H}
{Heller}, C.~H. 1995, \apj, 455, 252

\bibitem[{{Hillenbrand}(1997)}]{1997AJ....113.1733H}
{Hillenbrand}, L.~A. 1997, \aj, 113, 1733

\bibitem[{{Hillenbrand}(2002)}]{2002astro.ph.10520H}
{Hillenbrand}, L.~A. 2002, ArXiv Astrophysics e-prints

\bibitem[{{Hillenbrand}(2005)}]{2005astro.ph.11083H}
{Hillenbrand}, L.~A. 2005, ArXiv Astrophysics e-prints

\bibitem[{{Hillenbrand} \& {Carpenter}(2000)}]{2000ApJ...540..236H}
{Hillenbrand}, L.~A. \& {Carpenter}, J.~M. 2000, \apj, 540, 236

\bibitem[{{Hillenbrand} \& {Hartmann}(1998)}]{1998ApJ...492..540H}
{Hillenbrand}, L.~A. \& {Hartmann}, L.~W. 1998, \apj, 492, 540

\bibitem[{{Hillenbrand} {et~al.}(1998){Hillenbrand}, {Strom}, {Calvet},
  {Merrill}, {Gatley}, {Makidon}, {Meyer}, \&
  {Skrutskie}}]{1998AJ....116.1816H}
{Hillenbrand}, L.~A., {Strom}, S.~E., {Calvet}, N., {et~al.} 1998, \aj, 116,
  1816

\bibitem[{{Huff} \& {Stahler}(2006)}]{2006ApJ...644..355H}
{Huff}, E.~M. \& {Stahler}, S.~W. 2006, \apj, 644, 355

\bibitem[{{Jeffries}(2007)}]{2007MNRAS.376.1109J}
{Jeffries}, R.~D. 2007, \mnras, 376, 1109

\bibitem[{{Jones} \& {Walker}(1988)}]{1988AJ.....95.1755J}
{Jones}, B.~F. \& {Walker}, M.~F. 1988, \aj, 95, 1755

\bibitem[{{Kouwenhoven} {et~al.}(2007){Kouwenhoven}, {Brown}, {Portegies
  Zwart}, \& {Kaper}}]{2007A&A...474...77K}
{Kouwenhoven}, M.~B.~N., {Brown}, A.~G.~A., {Portegies Zwart}, S.~F., \&
  {Kaper}, L. 2007, \aap, 474, 77

\bibitem[{{Kroupa}(1995{\natexlab{a}})}]{1995MNRAS.277.1491K}
{Kroupa}, P. 1995{\natexlab{a}}, \mnras, 277, 1491

\bibitem[{{Kroupa}(1995{\natexlab{b}})}]{1995MNRAS.277.1507K}
{Kroupa}, P. 1995{\natexlab{b}}, \mnras, 277, 1507

\bibitem[{{Kroupa}(2002)}]{2002Sci...295...82K}
{Kroupa}, P. 2002, Science, 295, 82

\bibitem[{{Lada} \& {Lada}(2003)}]{2003ARA&A..41...57L}
{Lada}, C.~J. \& {Lada}, E.~A. 2003, \araa, 41, 57

\bibitem[{{Lada} {et~al.}(2000){Lada}, {Muench}, {Haisch}, {Lada}, {Alves},
  {Tollestrup}, \& {Willner}}]{2000AJ....120.3162L}
{Lada}, C.~J., {Muench}, A.~A., {Haisch}, Jr., K.~E., {et~al.} 2000, \aj, 120,
  3162

\bibitem[{{Lada} {et~al.}(2004){Lada}, {Muench}, {Lada}, \&
  {Alves}}]{2004AJ....128.1254L}
{Lada}, C.~J., {Muench}, A.~A., {Lada}, E.~A., \& {Alves}, J.~F. 2004, \aj,
  128, 1254

\bibitem[{{Lin} {et~al.}(2006){Lin}, {Ohashi}, {Lim}, {Ho}, {Fukagawa}, \&
  {Tamura}}]{2006ApJ...645.1297L}
{Lin}, S.-Y., {Ohashi}, N., {Lim}, J., {et~al.} 2006, \apj, 645, 1297

\bibitem[{{Luhman} {et~al.}(2000){Luhman}, {Rieke}, {Young}, {Cotera}, {Chen},
  {Rieke}, {Schneider}, \& {Thompson}}]{2000ApJ...540.1016L}
{Luhman}, K.~L., {Rieke}, G.~H., {Young}, E.~T., {et~al.} 2000, \apj, 540, 1016

\bibitem[{{Malkov} \& {Zinnecker}(2001)}]{2001MNRAS.321..149M}
{Malkov}, O. \& {Zinnecker}, H. 2001, \mnras, 321, 149

\bibitem[{{Mathieu}(1994)}]{1994ARA&A..32..465M}
{Mathieu}, R.~D. 1994, \araa, 32, 465

\bibitem[{{Mayne} \& {Naylor}(2008)}]{2008arXiv0801.4085M}
{Mayne}, N.~J. \& {Naylor}, T. 2008, ArXiv e-prints, 801

\bibitem[{{McCaughrean} {et~al.}(2002){McCaughrean}, {Zinnecker}, {Andersen},
  {Meeus}, \& {Lodieu}}]{2002Msngr.109...28M}
{McCaughrean}, M., {Zinnecker}, H., {Andersen}, M., {Meeus}, G., \& {Lodieu},
  N. 2002, The Messenger, 109, 28

\bibitem[{{Menten} {et~al.}(2007){Menten}, {Reid}, {Forbrich}, \&
  {Brunthaler}}]{2007A&A...474..515M}
{Menten}, K.~M., {Reid}, M.~J., {Forbrich}, J., \& {Brunthaler}, A. 2007, \aap,
  474, 515

\bibitem[{{Meyer} {et~al.}(1997){Meyer}, {Calvet}, \&
  {Hillenbrand}}]{1997AJ....114..288M}
{Meyer}, M.~R., {Calvet}, N., \& {Hillenbrand}, L.~A. 1997, \aj, 114, 288

\bibitem[{{Moeckel} \& {Bally}(2006)}]{2006ApJ...653..437M}
{Moeckel}, N. \& {Bally}, J. 2006, \apj, 653, 437

\bibitem[{{Moeckel} \& {Bally}(2007{\natexlab{a}})}]{2007ApJ...661L.183M}
{Moeckel}, N. \& {Bally}, J. 2007{\natexlab{a}}, \apjl, 661, L183

\bibitem[{{Moeckel} \& {Bally}(2007{\natexlab{b}})}]{2007ApJ...656..275M}
{Moeckel}, N. \& {Bally}, J. 2007{\natexlab{b}}, \apj, 656, 275

\bibitem[{{Muench} {et~al.}(2002){Muench}, {Lada}, {Lada}, \&
  {Alves}}]{2002ApJ...573..366M}
{Muench}, A.~A., {Lada}, E.~A., {Lada}, C.~J., \& {Alves}, J. 2002, \apj, 573,
  366

\bibitem[{{Olczak} {et~al.}(2006){Olczak}, {Pfalzner}, \&
  {Spurzem}}]{2006ApJ...642.1140O}
{Olczak}, C., {Pfalzner}, S., \& {Spurzem}, R. 2006, \apj, 642, 1140

\bibitem[{{{\"O}pik}(1924)}]{1924TarObs..6...25S}
{{\"O}pik}, E. 1924, Tartu Obs. Publ., 25, No. 6

\bibitem[{{Padgett} {et~al.}(1997){Padgett}, {Strom}, \&
  {Ghez}}]{1997ApJ...477..705P}
{Padgett}, D.~L., {Strom}, S.~E., \& {Ghez}, A. 1997, \apj, 477, 705

\bibitem[{{Palla} \& {Stahler}(2000)}]{2000ApJ...540..255P}
{Palla}, F. \& {Stahler}, S.~W. 2000, \apj, 540, 255

\bibitem[{{Pfalzner}(2006)}]{2006ApJ...652L.129P}
{Pfalzner}, S. 2006, \apjl, 652, L129

\bibitem[{{Pfalzner} \& {Olczak}(2007{\natexlab{a}})}]{2007A&A...462..193P}
{Pfalzner}, S. \& {Olczak}, C. 2007{\natexlab{a}}, \aap, 462, 193

\bibitem[{{Pfalzner} \& {Olczak}(2007{\natexlab{b}})}]{2007A&A...475..875P}
{Pfalzner}, S. \& {Olczak}, C. 2007{\natexlab{b}}, \aap, 475, 875

\bibitem[{{Pfalzner} {et~al.}(2006){Pfalzner}, {Olczak}, \&
  {Eckart}}]{2006A&A...454..811P}
{Pfalzner}, S., {Olczak}, C., \& {Eckart}, A. 2006, \aap, 454, 811

\bibitem[{{Reipurth} {et~al.}(2007){Reipurth}, {Guimar{\~a}es}, {Connelley}, \&
  {Bally}}]{2007AJ....134.2272R}
{Reipurth}, B., {Guimar{\~a}es}, M.~M., {Connelley}, M.~S., \& {Bally}, J.
  2007, \aj, 134, 2272

\bibitem[{{Rucinski}(2001)}]{2001AJ....122.1007R}
{Rucinski}, S.~M. 2001, \aj, 122, 1007

\bibitem[{{Scally} \& {Clarke}(2001)}]{2001MNRAS.325..449S}
{Scally}, A. \& {Clarke}, C. 2001, \mnras, 325, 449

\bibitem[{{Scally} {et~al.}(2005){Scally}, {Clarke}, \&
  {McCaughrean}}]{2005MNRAS.358..742S}
{Scally}, A., {Clarke}, C., \& {McCaughrean}, M.~J. 2005, \mnras, 358, 742

\bibitem[{{Shatsky} \& {Tokovinin}(2002)}]{2002A&A...382...92S}
{Shatsky}, N. \& {Tokovinin}, A. 2002, \aap, 382, 92

\bibitem[{{Sicilia-Aguilar} {et~al.}(2006){Sicilia-Aguilar}, {Hartmann},
  {Calvet}, {Megeath}, {Muzerolle}, {Allen}, {D'Alessio}, {Mer{\'{\i}}n},
  {Stauffer}, {Young}, \& {Lada}}]{2006ApJ...638..897S}
{Sicilia-Aguilar}, A., {Hartmann}, L., {Calvet}, N., {et~al.} 2006, \apj, 638,
  897

\bibitem[{{Sicilia-Aguilar} {et~al.}(2005){Sicilia-Aguilar}, {Hartmann},
  {Szentgyorgyi}, {Fabricant}, {F{\H u}r{\'e}sz}, {Roll}, {Conroy}, {Calvet},
  {Tokarz}, \& {Hern{\'a}ndez}}]{2005AJ....129..363S}
{Sicilia-Aguilar}, A., {Hartmann}, L.~W., {Szentgyorgyi}, A.~H., {et~al.} 2005,
  \aj, 129, 363

\bibitem[{{Slesnick} {et~al.}(2004){Slesnick}, {Hillenbrand}, \&
  {Carpenter}}]{2004ApJ...610.1045S}
{Slesnick}, C.~L., {Hillenbrand}, L.~A., \& {Carpenter}, J.~M. 2004, \apj, 610,
  1045

\bibitem[{{Stassun} {et~al.}(1999){Stassun}, {Mathieu}, {Mazeh}, \&
  {Vrba}}]{1999AJ....117.2941S}
{Stassun}, K.~G., {Mathieu}, R.~D., {Mazeh}, T., \& {Vrba}, F.~J. 1999, \aj,
  117, 2941

\bibitem[{{Sterzik} \& {Durisen}(1995)}]{1995A&A...304L...9S}
{Sterzik}, M.~F. \& {Durisen}, R.~H. 1995, \aap, 304, L9+

\bibitem[{{Tan}(2004)}]{2004ApJ...607L..47T}
{Tan}, J.~C. 2004, \apjl, 607, L47

\bibitem[{{Trimble}(1990)}]{1990MNRAS.242...79T}
{Trimble}, V. 1990, \mnras, 242, 79

\bibitem[{{Valtonen}(2004)}]{2004RMxAC..21..147V}
{Valtonen}, M. 2004, in Revista Mexicana de Astronomia y Astrofisica Conference
  Series, Vol.~21, Revista Mexicana de Astronomia y Astrofisica Conference
  Series, ed. C.~{Allen} \& C.~{Scarfe}, 147--151

\bibitem[{{Valtonen} \& {Mikkola}(1991)}]{1991ARA&A..29....9V}
{Valtonen}, M. \& {Mikkola}, S. 1991, \araa, 29, 9

\bibitem[{{Valtonen} {et~al.}(2005){Valtonen}, {Myll{\"a}ri}, {Orlov}, \&
  {Rubinov}}]{2005MNRAS.364...91V}
{Valtonen}, M., {Myll{\"a}ri}, A., {Orlov}, V., \& {Rubinov}, A. 2005, \mnras,
  364, 91

\bibitem[{{Valtonen}(1997)}]{1997ApJ...485..785V}
{Valtonen}, M.~J. 1997, \apj, 485, 785

\bibitem[{{van Altena} {et~al.}(1988){van Altena}, {Lee}, {Lee}, {Lu}, \&
  {Upgren}}]{1988AJ.....95.1744V}
{van Altena}, W.~F., {Lee}, J.~T., {Lee}, J.-F., {Lu}, P.~K., \& {Upgren},
  A.~R. 1988, \aj, 95, 1744

\bibitem[{{Wilson} {et~al.}(1997){Wilson}, {Filges}, {Codella}, {Reich}, \&
  {Reich}}]{1997A&A...327.1177W}
{Wilson}, T.~L., {Filges}, L., {Codella}, C., {Reich}, W., \& {Reich}, P. 1997,
  \aap, 327, 1177

\bibitem[{{Zinnecker} \& {Yorke}(2007)}]{2007ARA&A..45..481Z}
{Zinnecker}, H. \& {Yorke}, H.~W. 2007, \araa, 45, 481

\end{thebibliography}

%

\appendix

\section{Observability of tidal tails due to star-disc encounters in the ONC}
\label{app:enc_rate}
The relevant physical quantity that determines the prominence of tidal tails due to an encounter is the change of angular momentum in the disc.
According to \citet{2007A&A...462..193P}, we assume that a fractional angular momentum loss (AML) of at least 10\,\% is required to form observationally
detectable tidal tails \citep[see also Fig.~9 and 10 of][]{2007A&A...462..193P}.

To have an estimate of the rate of encounters in the ONC in which the AML is at least 10\,\%, we consider an encounter of a star with mass
$m$\,=\,$0.5$\,{\mbox{M$_{\odot}$}}, which corresponds to the mean stellar mass in the ONC (see \S\ref{sec:onc_dyn}). A star of that mass is thought to be surrounded by a
plotoplanetary disc of typical size $r_{\rm{d}}$\,=\,100\,AU. To be on the safe side, we want to estimate the upper limit of the encounter rate and further
assume the encounter partner to be the highest mass star of the ONC, $M$\,=\,50\,{\mbox{M$_{\odot}$}}. Referring to Table~1 of \citet{2007A&A...462..193P}, the specified
minimum AML requires an encounter at a relative periastron $r_{\rm{p}}/r_{\rm{d}}$\,$\approx$\,10, or a periastron $r_{\rm{p}}$\,$\approx$\,1000\,AU. Moreover,
we assume that the encounter occurred in the Trapezium Cluster (TC), the dense central part of the ONC, where it is most probable \citep[see Fig.~4
of][]{2006A&A...454..811P}. The number of stars, the density and the velocity dispersion of the TC are $N_{\rm{TC}}$\,$\approx$\,750, $n_{\rm{TC}}$\,$\approx$\,$10^3$\,pc$^{-3}$, and {\mbox{$\sigma_{\rm{1D}}^{\rm{JW}}$}}\,=\,2.5\,{\mbox{km\,s$^{-1}$}} (see \S\ref{sec:onc_dyn}).

The time scale for encounters of the assumed type is given by Eq. (\ref{eqn:tcoll}); substituting $n$\,=\,$n_{\rm{TC}}$, $\sigma$\,=\,{\mbox{$\sigma_{\rm{1D}}^{\rm{JW}}$}},
$r_{\ast}$\,=\,$r_{\rm{p}}$, and $m_{\ast}$\,=\,$M$, we arrive at $t_{\rm{enc}}$\,$\approx$\,2\,$\cdot$\,$10^5$\,yr. With a dissipation time scale of the
tidal tails of $t_{\rm{diss}}$\,$\lesssim$\,1000\,yr, the probability of a detection is roughly $P_{\rm{obs}}$\,$\lesssim$\,$t_{\rm{diss}} /
t_{\rm{enc}}$\,$\approx$\,5\,$\cdot$\,$10^{-3}$. With the number of stars in the TC, $N_{\rm{TC}}$, we expect at most
$N_{\rm{obs}}$\,$\lesssim$\,$P_{\rm{obs}} N_{\rm{TC}}$\,$\approx$\,4 stars to be accompanied by tidal tails that could be observed at the current time.

\section{Estimate of the mean uncertainty of stellar ages}
\label{app:age_errors}
Since individual errors of the derived stellar ages are not provided by \citet{1997AJ....113.1733H}, we estimate the mean error from the quoted
observational and theoretical uncertainties and the constructed HR diagram. The uncertainties of the derived luminosities, $\log{(L_\ast/L_\odot)}
\lesssim 0.2$, translate into an age uncertainty of $\sim$0.2-0.4\,dex for stars with masses 1.0-0.1\,{\mbox{M$_{\odot}$}}. Uncertainties of the derived effective
temperatures, $\log{T_\mathrm{eff}} \lesssim 0.02$, translate into an age uncertainty of typically $\sim$0.3\,dex, but can be as large as $\sim$1\,dex for
stars with mass $\lesssim$0.1\,{\mbox{M$_{\odot}$}}. Further uncertainties of the derived ages are introduced due to differences between different pre-main sequence
evolutionary tracks, which can be as large as 0.6\,dex. Assuming a typical uncertainty of 0.3\,dex due to uncertainties from luminosity, effective
temperature and evolutionary tracks we estimate a mean uncertainty of stellar ages of $\sim$0.5\,dex.

\section{Minimum velocity for unperturbed escape of stars in the ONC}
\label{app:enc_escape}
High-velocity stars that have been generated due to a close triple encounter in the cluster centre are expected to leave the cluster without significant
perturbation. This is due to the fact that (a) the fractional change of the velocity $\upsilon$ of a high-velocity star is less than 10\,\% unless the
impact parameter is not lower than 100\,AU \citep[][Eq.~4-8]{1987gady.book.....B}, and (b) only a small fraction of stars experiences more than one
encounter closer than 100\,AU in one crossing time of the ONC \citep{2001MNRAS.325..449S,2006ApJ...642.1140O}. {Alternatively, one can evaluate the
    collision time scale of the escaper and a binary with semi-major axis $a$\,$\approx$\,100\,AU in the ONC \citep[][Eq.~8-123]{1987gady.book.....B},
\begin{eqnarray}
\label{eqn:tcoll}
t_\mathrm{coll} = \left[ 16 \sqrt{\pi} n \sigma r_\ast^2 \left(1 + \frac{Gm_\ast}{2\sigma^2r_\ast}\right) \right]^{-1} \approx 10\,\mathrm{Myr},
\end{eqnarray}
where the gravitational focusing of a 10\,{\mbox{M$_{\odot}$}} binary in the dense Trapezium Cluster has been considered in order to obtain a robust lower limit on
$t_{\rm{coll}}$. We have used $r_\ast$\,=\,$a$, $n$\,=\,$ n_{\rm{TC}}$\,$\approx$\,$10^3$\,pc$^{-3}$ and $\sigma$\, =\,{\mbox{$\sigma_{\rm{3D}}^{\rm{JW}}$}}\,=\,$\sqrt{3}
{\mbox{$\sigma_{\rm{1D}}^{\rm{JW}}$}}$\,=\,4.3\,{\mbox{km\,s$^{-1}$}} (see~\S\ref{sec:onc_dyn}). Since the collision time scale is much longer than the time to reach the cluster outskirts,
$t_\mathrm{coll} \gtrsim 10\,\mathrm{Myr} \gg 0.2\,\mathrm{Myr} \gtrsim t_\mathrm{esc}$,} it is valid to assume that stars with velocities
$\upsilon$\,$\ge$\,3{\mbox{$\sigma_{\rm{3D}}^{\rm{JW}}$}} are effectively unperturbed before they escape from the cluster.

\section{Maximum velocity of a star ejected from a bound triple system}
\label{app:max_vel_triple_escaper}
{We analyze a triple system after the ejection event forming a configuration of a binary and an escaping body \citep[cf.][]{2005MNRAS.364...91V}. The
total energy of the system is then
\begin{eqnarray}
\label{eq:total_energy_triple}
E_0 = \frac{1}{2}m\dot{r}_\mathrm{s}^2 - G\frac{m_\mathrm{B}m_\mathrm{s}}{r_\mathrm{s}} + \frac{1}{2}M\dot{r}^2 - G\frac{m_\mathrm{a}m_\mathrm{b}}{r},
\end{eqnarray}
where $r_\mathrm{s}$ is the separation of the third body relative to the barycentre of the binary, $r$ the separation of the binary components,
$m_\mathrm{s}$, $m_\mathrm{a}$ and $m_\mathrm{b}$ the masses of the escaper and the binary components, $m_\mathrm{B} = m_\mathrm{a} + m_\mathrm{b}$
the binary mass, and $M = m_\mathrm{a} m_\mathrm{b}/m_\mathrm{B}$ and $m = m_\mathrm{B} m_\mathrm{s}/(m_\mathrm{B} + m_\mathrm{s})$ the reduced masses.

When observing, the detection of the escaper will usually occur when the distance to the binary components greatly exceeds the size of the binary
system, $r_\mathrm{s} \gg r$. Thus we can neglect the second term in Eq.~(\ref{eq:total_energy_triple}). Moreover, in most cases the ejected body will be the
lowest mass component of the triple system, $m_\mathrm{s} \ll m_\mathrm{B}$ (see~\S\ref{sec:onc_enc}), hence Eq.~(\ref{eq:total_energy_triple}) reduces to
\begin{eqnarray}
\label{eq:total_energy_triple_reduced}
E_0 = \frac{1}{2}m_\mathrm{s}\dot{r}_\mathrm{s}^2 + \frac{1}{2}M\dot{r}^2 - G\frac{m_\mathrm{a}m_\mathrm{b}}{r}
    = \frac{1}{2}m_\mathrm{s}\dot{r}_\mathrm{s}^2 - G\frac{m_\mathrm{a}m_\mathrm{b}}{2a},
\end{eqnarray}
where the last term denotes the total binary energy in the general case of an elliptical orbit with semi-major axis $a$.

If we use the same parameters for the triple system as in \S\ref{sec:onc_enc} and assume a nearly equal-mass binary, $m_\mathrm{s}$\,=\,0.25\,{\mbox{M$_{\odot}$}},
$2a$\,=\,100\,AU, $m_\mathrm{a}m_\mathrm{b}$\,$\approx$\,4\,{\mbox{M$_{\odot}$}}, we find from Eq.~(\ref{eq:total_energy_triple_reduced}) the maximum
velocity of the escaper from a triple with negative total energy, $\upsilon_\mathrm{s} = \dot{r}_\mathrm{s} \lesssim 18\,{\mbox{km\,s$^{-1}$}}$.
In fact, our simulations show that the binary that generates a high-velocity star never is less massive then 10\,{\mbox{M$_{\odot}$}} and usually exceeds 20\,{\mbox{M$_{\odot}$}},
and the mass ratio is never below 1/6 and usually about 1/3. Using these parameters, we find $\upsilon_\mathrm{s}$\,$\lesssim$\,30-80\,{\mbox{km\,s$^{-1}$}}.}

%

\listofobjects

\end{document}